\documentclass{pasa}%

\usepackage{natbib}
\usepackage{hyperref}
\usepackage{epstopdf}
\usepackage{float}
\usepackage{wrapfig}
\usepackage{graphicx}
\usepackage{enumerate}  
\usepackage[normalem]{ulem}
\usepackage{mathptmx}
\usepackage{booktabs}
\usepackage[table,xcdraw]{xcolor}
\usepackage{multirow}
\usepackage{rotating,tabularx}

\definecolor{armygreen}{rgb}{0.29, 0.33, 0.13}
\definecolor{english}{rgb}{0.0, 0.5, 0.0}
\definecolor{airforceblue}{rgb}{0.3, 0.3, 0.8}
\definecolor{bgrose}{rgb}{0.28, 0.02, 0.03}
\definecolor{applegreen}{rgb}{0.55, 0.71, 0.00}
\definecolor{olivedrab}{rgb}{.25, 0.50, 0.2}
\definecolor{orange}{rgb}{0.45, 0.26, 0.0}
\definecolor{blue2}{rgb}{0.3, 0.1, 0.8}

\def\msun{M$_\odot$}
\def\lsun{L$_\odot$}

\newcommand{\chem}[1]{$^{#1}$}
\newcommand{\iso}[1]{$^{#1}$}

\title[Fates of the oldest intermediate-mass stars]{Primordial to Extremely Metal-Poor AGB and Super-AGB Stars: White Dwarf or Supernova progenitors?}

\author[Gil-Pons et al.]{Pilar Gil-Pons$^{1,2}$, Carolyn L. Doherty$^{3,4}$, Jordi L. Guti\'errez$^{1,2}$, Lionel Siess$^{5}$, Simon W. Campbell$^4$,\\ Herbert B. Lau \and John C. Lattanzio$^4$ \\
\affil{$^1$Polytechnical University of Catalonia, Barcelona, Spain}%
\affil{$^2$Institut d'Estudis Espacials de Catalunya, Barcelona, Spain}
\affil{$^3$Konkoly Observatory, Hungarian Academy of Sciences, 1121 Budapest }%
\affil{$^4$Monash Centre for Astrophysics, School of Physics and Astronomy, Monash University, Australia}%
\affil{$^5$Institut d'Astronomie et d'Astrophysique, Universit\'e Libre de Bruxelles, ULB, Belgium}}%

\jid{PASA}
\doi{10.1017/pas.\the\year}
\jyear{\the\year}

\usepackage{aas_macros}
\usepackage{hyperref} 
\hypersetup{colorlinks,citecolor=blue,linkcolor=blue,urlcolor=blue}
\begin{document}%

\begin{abstract}

Getting a better understanding of the evolution and nucleosynthetic yields of the most metal-poor stars ($Z\lesssim 10^{-5}$) is critical because they are part of the big picture of the history of the primitive Universe. Yet many of the remaining unknowns of stellar evolution lie in the birth, life, and death of these objects.
We review stellar evolution of intermediate-mass $Z \leq 10^{-5}$ models existing in the literature, with a particular focus on the problem of their final fates. We emphasize the importance of the mixing episodes between the stellar envelope and the nuclearly processed core, which occur after stars exhaust their central He (second dredge-up and dredge-out episodes). The depth and efficiency of these episodes are critical to determine the mass limits for the formation of electron-capture supernovae (EC-SNe). Our knowledge of these phenomena is not complete because they are strongly affected by the choice of input physics.
These uncertainties affect stars in all mass and metallicity ranges. However, difficulties in calibration pose additional challenges in the case of the most metal-poor stars.
We also consider the alternative SN~I1/2 channel to form supernovae out of the most metal-poor intermediate mass objects. In this case, it is critical to understand the thermally-pulsing AGB evolution until the late stages. Efficient second dredge-up and, later, third dredge-up episodes could be able to pollute stellar envelopes enough for the stars to undergo thermal pulses in a way very similar to that of higher initial $Z$ objects. Inefficient second and/or third dredge-up may leave an almost pristine envelope, unable to sustain strong stellar winds. This may allow the H-exhausted core to grow to the Chandrasekhar mass before the envelope is completely lost, and thus let the star explode as a SN~I1/2. After reviewing the information available on these two possible channels for the formation of supernovae, we discuss existing nucleosynthetic yields of stars of metallicity $Z \leq 10^{-5}$, and present an example of nucleosynthetic calculations for a thermally-pulsing Super-AGB star of $Z = 10^{-5}$. We compare theoretical predictions with observations of the lowest [Fe/H] objects detected. The review closes by discussing current open questions as well as possible fruitful avenues for future research.
\end{abstract}
\maketitle%

\section{INTRODUCTION}\label{sec:intro}

The evolution and nucleosynthesis of the most metal-poor stars and, in particular, the determination of the mass thresholds for the formation of supernovae at the lowest metallicity regimes, hold some of the clues to understanding the formation and early chemical evolution of galaxies.

According to the $\Lambda-$Cold Dark Matter model, the current standard model of Big Bang cosmology, the first stars\footnote{Given its origin and composition the first stars have also been named primordial, metal-free, hydrogen-helium stars, or population III (Pop III) stars.} formed at redshift $z \sim$ 20-30, just a few hundred million years after the Big-Bang, in $\sim 10^6$~\msun{} minihaloes where atomic gas and traces of ${\rm H_2}$ could efficiently condense and radiatively cool. This theory was presented by \cite{couchman1986} and \cite{tegmark1997}, although the interest in the evolution of metal-free stars dates from more than two decades earlier. \cite{ezer1961} computed pure hydrogen zero-age main sequence models over a wide range of masses. \cite{truran1971} proposed that the first stars in the universe were the direct nucleosynthetic heirs of the Big-Bang. This origin determined their pristine composition, consisting of H, He and trace amounts of light elements.

During the 1970s the interest in the evolution of metal-free and very metal-poor stars was consolidated, and it has continued to the present day. Simultaneously, the study of primordial star formation and of the primitive initial mass function (IMF) developed. The debate on the possibility of occurrence of non-massive metal-free stars, and on the actual shape of the ancient IMF began. High resolution multi-dimensional hydrodynamical calculations have recently confirmed the possibility of forming primordial low-mass stars (see, for instance, \cite{sus14} and references therein). Nevertheless the concept of critical metallicity \citep{bromm2001}, which refers to the minimum metal content required for the formation of low-mass stars, seems to be observationally supported \citep{frebel2007}, and thus the debate over the existence of low-mass primordial stars is not over yet.

Given the uncertainties in the IMF for the most metal-poor stars, and the lack of observational constraints, we 
must face the uncertainty of their existence, although so must those studying hyper-massive stars \citep{heger2001}. Metal-poor models are further hampered by many unknowns, mostly related to stellar mixing, the location of convective boundaries, and mass-loss rates due to stellar winds. These uncertainties also affect stellar modelling at higher $Z$ (see, for instance, the discussion in \citealt{doherty2017} and references therein), although in such cases calibration by comparison with observations is 
more often feasible and some restrictions on input physics can be obtained. This is not the case in the most metal-poor regime because of different reasons. First, the possibility of comparing with observations is limited because of the relatively small sample of detected objects in the most metal-poor regime. At present only $\sim$ 10 stars are known to have metallicity [Fe/H]\footnote{$\rm [Fe/H]=\log{(N_{Fe}/N_H)_\star}-\log{(N_{Fe}/N_H)_{\odot}}$, where the subscript $\ast$ refers to the considered star, and $N$ is the number density.} $ < -4.5$ (\citealt{starkenburg2017}, \citealt{aguado2018}, and \citealt{bonifacio2018} and references therein). The record is held by the star detected by \cite{keller2014}, with ${\rm [Fe/H] < -7.1}$. 
As metallicity increases, so does the number of observed stars. According to
the SAGA database (\citealt{suda2008}, \citealt{suda2011}, \citealt{yamada2013}, \citealt{suda2017a}), there are $\sim$ 500 stars with [Fe/H] $< -3$.
Second, even the most metal-poor stars detected may be the descendents of not one but a few approximately coeval objects. Their surface abundances may have suffered some degree of pollution due to internal processes such as dredge-up episodes,  and accretion from the interstellar medium. Finally, as will be reviewed in this work, computation of the evolution of the most metal-poor stars is very demanding. Low-mass stars experience violent flashes which put hydrostatic codes at the limit of their performance (\citealt{picardi2004}, \citealt{campbell2008}, and \citealt{woodward2015}); more massive objects can experience  thousands of thermal pulses \citep{lau2008,gilpons2013} and not only their detailed nucleosynthetic yields but even their fates as white dwarfs or supernovae is, at present, uncertain for relatively wide ranges of initial masses and metallicities. 

The evolution of stars of metallicity $Z \gtrsim 10^{-4}-10^{-3}$ has been extensively studied and is relatively well understood (see, for instance \citealt{iben2012}). Their fate depends primarily on their mass, but the initial composition, input physics or the presence of a companion star can also play a crucial role and modify their fate. Traditionally single stars with initial mass  $\rm M_{ZAMS}$ $\lesssim$ 7--10~\msun{} (depending on the metallicity) will develop a degenerate core 
and end their lives as white dwarfs. The more massive counterparts on the other end will go through all nuclear burning stages and explode as core collapse supernovae. However, in between these two recognised stellar components, there is a very narrow mass range of 0.2--0.5~\msun{} width beyond the maximum mass for the formation of white dwarfs where stars are likely to evolve as electron-capture supernovae (EC-SNe, EC-SN for the singular). 
These explosions are triggered by electron captures on $^{24}$Mg and $^{20}$Ne in the degenerate ONe core. EC-SNe have attracted interest in the 1980's (\citealt{miy80}, \citealt{nomoto1984} and \citealt{nomoto1987}), and models have been subsequently improved. More realistic EC-SN progenitors, including the evolution from the main sequence, with updated input physics, and closer to the time of the explosion have been presented since then 
\citep{ritossa1999,jones2016a}. 

Intermediate-mass stars can be defined as those of mass high enough to avoid a core-He flash, but not massive enough to end their lives as core-collapse SNe. They become white dwarfs when they are able to lose their envelopes by stellar winds before their cores reach the Chandrasekhar mass, $\rm M_{Ch}$. 
If some mechanism prevents envelope ejection before the core reaches $\rm M_{Ch}$, a supernova explosion would ensue. 
This type of supernova (in a metallicity-independent context) was  first proposed by \cite{arnett1969} and
later named SNI1/2 by \cite{ibe83}, after considering that the explosion mechanism should be similar to a that of a thermonuclear Type Ia SN, but that these objects should show hydrogen in their spectra, like a Type-II SN. 
According to \cite{ibe83} SNI1/2 explosions could be expected at least for the most massive AGB stars, 
which experienced C ignition before their cores were reduced to masses below $\rm M_{Ch}$. However 
detailed evolutionary calculations (see \citealt{siess2010} and references therein) showed that this supernova mechanism was prevented by the ejection of the stellar envelope (through winds), before the core reached $\rm M_{Ch}$. Interest in SN~I1/2 grew again in the 2000s in the context of the evolution of primordial stars with very weak stellar winds. The possibility that they could have existed in the primitive universe was discussed first in \cite{zij04} and later in \cite{gilpons2007} and \cite{lau2008}.
Note that, as happens for higher metallicity stars, the occurrence of metal-poor intermediate-mass stars in close binary systems may drastically alter their evolution and fates.

Gaining insight into stellar evolution at the extremely-metal-poor regime ($\rm [Fe/H] \lesssim -3$ or $Z\lesssim 10^{-5}$, assuming scaled solar composition) represents a small but nevertheless potentially important part in the formidable problem of understanding the primitive Universe. It involves, besides stellar evolution and nucleosynthesis, additional inputs from different fields of astrophysics.
Cosmological and star formation theories should be considered, as well as interstellar medium physics, thermodynamical and chemical evolution, and galaxy formation theories (see, for instance, the review by \citealt{karlsson2013}).

Increasingly powerful computational resources enable us to construct refined models, and investigate a much more extended range of 
possible input physics. The huge increase in observational data of metal-poor stars coming from big surveys, such as the HK objective-prism survey \citep{beers1992}, the Hamburg-ESO survey \citep{christlieb2002}, SkyMapper \citep{keller2007}, the Sloan Extension for Galactic Understanding and Exploration, SEGUE \citep{yanny2009}, and the  Large Sky Area Multi-Object Fibre Spectroscopic Telescope, LAMOST \citep{cui2012}, will be further expanded with the new wide-field multi-object spectrograph for the William Herschel Telescope, WEAVE \citep{dalton2012}, the PRISTINE survey \citep{starkenburg14}, and, specially, with the James Webb Space Telescope \citep{zackrisson2011}. They will provide us with a wealth of information about the elusive [Fe/H] $\leq -4.5$ ($Z \lesssim 5\times 10^{-7}$) stars, to which the findings of the computational models described in this work relate. 

In the present work we compile and discuss our current knowledge of the evolution and fates of single intermediate-mass stars between primordial metallicity and $Z=10^{-5}$. For the sake of providing context, we also summarize the successes and problems of low- and high-mass stellar models in the interpretation of observations of metal-poor stars. This document is structured as follows. Section \ref{sec:history} reviews the history of the understanding of primordial star formation, and of stellar evolution at the lowest metallicities. Section \ref{sec:evolution} summarizes the evolution of intermediate-mass stars in the considered metallicity regime. Section \ref{sec:uncertainties} delves into the main uncertainties which affect our knowledge of these stars. Section \ref{sec:fates} is devoted to analysis of their final fates, considering different input physics. 
Section \ref{sec:observations} summarises the main features of the most metal-poor stars detected. Section  \ref{sec:nucleosynthesis} describes the nucleosynthesis of intermediate-mass stars of $Z \leq 10^{-5}$, and relates it to observational evidence introduced in Section \ref{sec:observations}. In the last section, the results presented in this review are discussed, and possible future lines of work are outlined.

The following nomenclature is used in the present manuscript. 
Unless otherwise stated, metallicity $Z$ is the total mass fraction of metals, meaning all species other than H and He. 
Metallicity may also be expressed by referring to solar values, such as via [Fe/H], according to the standard expression given in Footnote $2$. 
{\it Extremely metal-poor (EMP)} stars in this work refer to those whose metallicity $Z \leq 10^{-5}$.  Note that the standard definition of EMP corresponds to stars with [Fe/H]$ < -3$ \citep{beers2005}. Using standard solar composition values (see \citet{asplund2006} and references therein), $Z \sim 10^{-5}$ is equivalent to $\rm [Fe/H] < -3$, except for a few 0.1 dex. However it should be noted that, given their origin either as primordial or descendants of primitive supernovae, EMP stars are not expected to have abundances that are simply scaled versions of the solar composition, and observations confirm this trend (see, for instance \citet{bonifacio2015}, \citet{keller2014}, \citet{yong2013}, or \citet{caffau2011}).  
The entire metallicity range from $Z \sim 10^{-5}$ ([Fe/H]$\sim -3$) down to $Z \sim 0$ is included in the expression {\it primordial to EMP stars}.
According to \cite{beers2005} {\it ultra metal-poor} and {\it hyper metal-poor} stars refer to stars with $\rm [Fe/H] < -4$ and $\rm [Fe/H] < -5$ respectively.
Primordial stars have been computed either using a strict zero metal-content, or considering $Z_\mathrm{ZAMS}\sim 10^{-10}$. This value is above the expected Big-Bang nucleosynthesis metallicity \citep{coc2004} but, as we will show in Section \ref{sec:evolution}, it still preserves the characteristics of primordial star evolution.
Note also that the intermediate-mass stars we analyse, although initially metal-poor, may evolve to become highly enriched in metals during their evolution. 
Strictly speaking it would be more correct to refer to them as ``iron-poor", but we will 
still call them metal-poor, following the more frequent nomenclature in the literature.

\section{THE NATURE OF ANCIENT STARS AND THE HISTORY OF THEIR MODELLING}\label{sec:history}

The first models of stars composed purely of H and He started appearing in the literature during the early 1970s. 
The evolution of the main central H- and He-burning stages in a wide range of masses, from the low to the massive cases, was computed by \cite{ezer1971}, \cite{ezer1972}, and shortly afterwards by \cite{cary1974}, and \cite{castellani1975}. \cite{wagner1974} undertook the first exploration of the behaviour of stars as a function of metallicity $Z$, and concluded that this behaviour became independent of $Z$ for values $Z\lesssim 10^{-6}$. \cite{dantona1982} were the first to report the existence of a helium flash in a low-mass primordial star. 

Understanding the first stars also involves understanding their formation process and the primitive IMF. \cite{yoneyama1972} concluded that, in the absence of metals, primordial clouds would lack the dust and heavy molecules able to provide the necessary cooling and fragmentation mechanisms which drive the formation of non-massive stars\footnote{In general gas clouds can be fragmented by the amplification of density fluctuations caused by gravitational and/or thermal instabilities. Significant thermal instabilities require efficient cooling, as may be caused by atomic fine line emissions, by molecules transitioning to rotational or vibrational states of lower energy, or heating of dust grains. More efficient cooling and thus lower gas cloud temperatures lower the Jean's mass and favour fragmentation.}. This result was in sharp contrast to the present observed IMF (\citealt{salpeter1955}, \citealt{scalo1979}, \citealt{kroupa2001}, and \citealt{chabrier2003}), that favours low mass stars. 
\cite{carlberg1981}, and \cite{palla1983} found that absorption in the ${\rm H_2}$ molecule could provide the necessary cooling to form low-mass primordial stars. Also on the basis of ${\rm H_2}$-cooling, \cite{yoshii1986} reported a primordial IMF that peaked at intermediate-mass values, between 4 and 10~\msun.  The latter results motivated interest in a further study of the evolution and nucleosynthesis of the late stages of low- and intermediate-mass stars (as well as massive), and a number of works dealing with the absence or existence of the thermally-pulsing Asymptotic Giant Branch (AGB) of primordial stars were published (\citealt{castellani1983} , \citealt{chieffi1984}, and \citealt{fujimoto1984}). Later works of \cite{omukai1998} also supported the possibility of forming low-mass primordial stars, and \cite{nakamura2001a} determined a bimodal primordial IMF peaked both at about 1 and 10~\msun.

The big picture of the nature of the first stars changed again after 3D hydrodynamical simulations of primordial star formation by \cite{abel1998,abel02} and \cite{bromm03}, who concluded that primordial stars had to be very massive ($\rm M_{ZAMS}\gtrsim 10^3$~\msun). Pair-Instability Supernova
models, triggered by the production of electron-positron pairs at high entropy and temperature 
(e.g. \citealt{umeda2002}, \citealt{woosley2017}),  and very energetic core-collapse supernovae or hypernovae (e.g.\citealt{nakamura2001b}, \citealt{nomoto2002} gained popularity as the first polluters of the primitive universe.

The effects of rotation and induced mixing on the early evolution of primordial to very low-metallicity massive stars were also investigated \cite[e.g.][]{ekstrom2008} and the associated nucleosynthetic yields presented by various groups (\citealt{woosley1995}, \citealt{umeda2002}, \citealt{chieffi2002}, \citealt{chieffi2004}, \citealt{kob06}, \citealt{heger2010}, \citealt{limongi2012}, and \citealt{takahashi2014}).
In the context of primordial massive star models, it is also important to consider the success of supernova yields in interpreting observations of metal-poor stars (\citealt{umeda2003}, \citealt{lim03}, \citealt{bonifacio2003}, \citealt{rya05}, \citealt{kob14}, and \citealt{tominaga2014}).

Despite the uncertainty of the existence of non-massive stars in the lowest $Z$ regime, many groups continued the study of their evolution (\citealt{hollowell1990}, \citealt{fujimoto2000}, \citealt{weiss2000}, \citealt{dominguez2000}, \citealt{chieffi2001},  \citealt{schlattl2001}, \citealt{siess2002}, \citealt{gilpons2005}, \citealt{gilpons2007}, \citealt{campbell2008}, and \citealt{lau2008}). The characteristics of the thermally-pulsing AGB and Super-AGB, the nucleosynthetic yields, and even the elusive final fates of some of these stars were outlined and debated.

Increasingly higher resolution simulations of star formation suggested that photoionisation and photoevaporation were able to halt mass-accretion onto metal-free protostars. As a consequence primordial stars of masses in the range 50--300~\msun{} were able to form (\citealt{mckee08}, and \citealt{bromm09}). 
Other simulations (\citealt{sta14}, \citealt{hir14}, \citealt{sus14}), with even higher resolution, opened the possibility of forming low- and intermediate-mass stars in primordial environments. Additionally, further fragmentation of circumstellar disks could result in binary or multiple stellar systems composed of low-mass objects \citep{clark2011}.
Yet, until recently, the preferred perspective among a large part of the scientific community was that Pop III stars were massive or very massive. Pop III refer to the first (metal-free) generation of stars. Pop II corresponds to subsequent generations, formed from metal-poor gas ejected by Pop III objects and their progeny. Pop I are young (metal-rich) stars.

\cite{omukai2000}, \cite{bromm2001}, and \cite{spaans2005} introduced the concept of critical metallicity 
to describe the minimum metal content in star-forming gas clouds which could allow the formation of low-mass (Pop II) stars. The transition from environments able to host the formation of Pop III to those able to host the formation of Pop II stars was determined by the occurrence of additional gas-cooling mechanisms: line-cooling \citep{bromm03}, which gave a critical metallicity $Z_{\rm crit} \sim 10^{-3.5}\:Z_{\odot}$, and dust-induced fragmentation (\citealt{schneider2010}, 
\citealt{dopcke2013}), which gave $Z_{\rm crit}$ values 2-3 orders of magnitude lower than the line-cooling mechanism. 

The line-cooling mechanism and thus the existence of a critical luminosity seems to be observationally supported \citep{frebel2007}, although the absence of detection of stars below a certain metallicity might be simply a consequence of their rarity and low luminosities, or due to pollution resulting from accretion of interstellar material \citep{komiya2015}. 
However, doubts were shed on the latter results by \cite{tanaka2017} and \cite{suzuki2018}. 
\cite{schneider12} proposed that the dust produced during the evolution of primordial massive stars and supernova explosions could induce the fragmentation required to form Pop II low-mass stars.

\section{EVOLUTION OF PRIMORDIAL TO EXTREMELY METAL-POOR INTERMEDIATE-MASS STARS}\label{sec:evolution}

\begin{table*}
\begin{center}
\begin{tabular}{ccccc}
\hline
\hline
${\rm M_{ZAMS}\,(\rm M_{\odot})}$ & $Z=10^{-10}$ & $Z=10^{-8}$ & $Z=10^{-6}$ & $Z=10^{-5}$ \\
\hline
3.0 & 227.8 & 229.6 & 236.4 & 246.1  \\
4.0 & 114.9 & 117.3 & 124.5 & 124.6  \\
5.0 & 68.9 & 71.3 & 77.1 & 77.6 \\
6.0 & 46.5 & 48.7 & 53.4 & 54.9 \\
7.0 & 34.0 & 36.2 & 40.1 & 41.0 \\
8.0 & 26.3 & 28.2 & 31.7 & 32.3 \\
9.0 & 21.5 & 23.1 & 26.0 & 26.4 \\
9.5 & 19.5 & 21.2 & 24.0 & 24.5 \\
\hline
\end{tabular}\\

\caption{
Times (in Myr) at the end of our calculations for selected EMP example models. Calculations were halted during  the later stages of the thermally-pulsing AGB or Super-AGB. 
}
\label{times}
\end{center}
\end{table*}

\begin{table*}
\begin{center}
\vspace{0.2cm}
$Z=10^{-10}$
\tabcolsep=0.15cm
\begin{tabular}{llllllllllll}\\
\hline
   & CHB & ${\rm CHB_{end}}$ &  & ${\rm CHeB_{begin}}$ 
   & ${\rm CHeB_{end}}$ 
   & &  & & & ${\rm CCB_{begin}}$ \\
\hline
\hline
${\rm M_{ZAMS}}$ & 
${\rm M_{cc}}$ & 
$X_{\rm c}(C)$ & $X_{\rm c}(O)$& 
${\rm M_{HexC}}$ & 
${\rm M_{HexC}}$ &
$X_{\rm HBS}(C)$ & $X_{\rm HBS}(N)$ & $X_{\rm HBS}(O)$ & 
$(C/O)_{\rm c}$ & ${\rm M_{Cign}}$ \\
\hline
3.0  &     0.25    &  1.6$\times 10^{-10}$  &   3.3$\times 10^{-11}$ &     0.44 &      0.63 &  
4.3$\times 10^{-10}$  & 1.6$\times 10^{-8}$ &   6.5$\times 10^{-11}$ &    1.36 & -- 
\\
4.0  &     0.36    &  1.0$\times 10^{-8}$ &     8.2$\times 10^{-11}$ &     0.53 &     0.78 &  
4.7$\times 10^{-10}$  & 1.9$\times 10^{-8}$ &   7.3$\times 10^{-11}$ &    1.34  & -- 
\\  
5.0 &     0.53    &   4.0$\times 10^{-7}$ &     1.9$\times 10^{-10}$ &     0.61 &    0.92 &  
5.1$\times 10^{-10}$  & 2.2$\times 10^{-8}$ &   7.9$\times 10^{-11}$ &    1.32  & --
\\
6.0 &     0.73    &   4.8$\times 10^{-6}$ &     1.2$\times 10^{-9}$ &     0.72 &    1.13 &  
5.4$\times 10^{-10}$  & 2.6$\times 10^{-8}$ &   8.2$\times 10^{-11}$ &    1.27  & --
\\      
7.0 &     0.79    &   2.8$\times 10^{-5}$ &     5.9$\times 10^{-9}$ &     0.78 &  1.29 &  
8.3$\times 10^{-10}$  & 3.1$\times 10^{-8}$ &   1.1$\times 10^{-10}$ &    1.26  & 0.57
\\ 
8.0 &     0.90     &  1.1$\times 10^{-4}$ &     1.3$\times 10^{-8}$ &     0.84 & 1.51 &  
9.1$\times 10^{-10}$  & 3.7$\times 10^{-8}$ &   4.9$\times 10^{-10}$ &    1.20  & 0.39
\\        
9.0 &     1.35    &  4.1$\times 10^{-4}$ &     6.8$\times 10^{-8}$ &     0.98 &  1.74 &  
1.2$\times 10^{-9}$   & 4.9$\times 10^{-8}$ &   1.6$\times 10^{-10}$ &    1.13  & 0.17
\\               
\hline
\end{tabular}
\end{center}
\vspace{1cm}
\begin{center}
$Z=10^{-5}$
\tabcolsep=0.15cm
\begin{tabular}{llllllllllll}\\
\hline
   & CHB & ${\rm CHB_{end}}$ &  & ${\rm CHeB_{begin}}$ 
   & ${\rm CHeB_{end}}$ 
   & &  & & & ${\rm CCB_{begin}}$ \\
\hline
\hline
${\rm M_{ZAMS}}$ & 
${\rm M_{cc}}$ & 
$X_{\rm c}(C)$ & $X_{\rm c}(O)$& 
${\rm M_{HexC}}$ & 
${\rm M_{HexC}}$ &
$X_{\rm HBS}(C)$ & $X_{\rm HBS}(N)$ & $X_{\rm HBS}(O)$ & 
$(C/O)_c$ & ${\rm M_{Cign}}$ \\
\hline
3.0 &     0.43     &  6.7$\times 10^{-8}$ &     1.3$\times 10^{-8}$ &     0.36 &      0.77 &  
1.1$\times 10^{-7}$  & 6.4$\times 10^{-6}$ &   5.9$\times 10^{-8}$ &    1.05 &  -- 
\\
4.0 &     0.86     &  8.4$\times 10^{-8}$ &     2.0$\times 10^{-8}$ &     0.48 &     0.96 &  
9.9$\times 10^{-8}$  & 6.7$\times 10^{-6}$ &   5.7$\times 10^{-8}$ &    0.99 &  --
\\  
5.0 &     1.25     &  9.4$\times 10^{-8}$ &     2.4$\times 10^{-8}$ &     0.62 &    1.16 &  
9.3$\times 10^{-8}$  & 6.8$\times 10^{-6}$ &   5.4$\times 10^{-8}$ &    1.03 &   --
\\
6.0 &     1.79     &  9.8$\times 10^{-8}$ &     2.8$\times 10^{-8}$ &  0.73   & 1.27   &  
8.7$\times 10^{-8}$  & 7.0$\times 10^{-6}$ &   5.2$\times 10^{-8}$ &  0.90   &  --
\\      
7.0 &     2.18     &  1.0$\times 10^{-7}$ &     3.2$\times 10^{-8}$ &     0.89 &  1.69 &  
8.5$\times 10^{-8}$  & 7.0$\times 10^{-6}$ &   5.2$\times 10^{-8}$ &  0.88 &   0.55
\\ 
8.0 &     2.55     &  1.3$\times 10^{-7}$ &     3.7$\times 10^{-8}$ &     1.32 & 1.94 &  
7.8$\times 10^{-8}$  & 7.0$\times 10^{-6}$ &   4.8$\times 10^{-8}$ &    0.96 &   0.26
\\        
9.0 &     3.09     &  1.4$\times 10^{-7}$ &    3.9$\times 10^{-8}$ &     1.90 &  2.24 &  
7.4$\times 10^{-8}$  & 7.1$\times 10^{-6}$ &   4.5$\times 10^{-12}$ &  0.98  &    0.02
\\               
\hline
\end{tabular}
\end{center}
\begin{center}
\caption{Relevant structure and composition parameters for the primordial and $Z=10^{-5}$ models. ${\rm M_{\rm cc}}$ represents the maximum size of the convective core during core H-burning (CHB). 
$X_{\rm c}(C)$ and $X_{\rm c}(O)$ are, respectively, the central abundances of C and O at the end of CHB. 
${\rm {M_{\rm HexC}}}$ in columns 5 and 6 refer to the size of the H-exhausted core at the beginning and at the end of core He-burning (CHeB).
$X_\mathrm{HBS}(C)$, $X_\mathrm{HBS}(N)$ and $X_\mathrm{HBS}(O)$ are abundances at the H-burning shell 
(at the mass point of its peak ${\rm ^{14}N}$ abundance) at the end of central He-burning. 
$(C/O)_c$ is the quotient of the central abundances of C and O at the same time. 
The last columns gives the mass point of C ignition. All masses are given in solar units.
Note that the end of CHB was taken when central H-abundance $X_{\rm c}(H) < 10^{-8}$.
The beginning of CHeB was taken when $\rm L_{\rm He}$= 100 L$_{\odot}$.
The end of CHeB was taken when central He-abundance $X_{\rm c}(He) < 10^{-8}$.
}\label{tab:refe1}
\end{center}
\end{table*}

\begin{table*}

\begin{center}
\vspace{0.5cm}
$Z=10^{-10}$\\
\tabcolsep=0.12cm
\begin{tabular}{ccccccccccc}\\
\hline
   & Bef. SDU & Aft. SDU &  &  &  & 1$^{\rm st}$ TP &  &  &  & \\
\hline
${\rm {M_{ZAMS}}}$ & 
${\rm {M_{HexC}}}$ & 
${\rm {M_{HexC}}}$ &
$X_{\rm s}(C)$ & $X_{\rm s}(N)$ & $X_{\rm s}(O)$   &
${\rm {M_{HeexC}}}$ & 
${\rm {M_{HexC}}}$ &
$X_{\rm s}(C)$ & $X_{\rm s}(N)$ & $X_{\rm s}(O)$
\\
\hline
\hline
3.0 &  1.02 &  1.02 & $1.7\times 10^{-11}$ &  $5.3\times 10^{-12}$ & $4.8\times 10^{-11}$ & 0.793 & 0.812 & 1.7$\times 10^{-11}$ &  $5.3\times 10^{-12}$ & $4.8\times 10^{-11}$ \\
4.0 &  0.87 &  0.87 & $3.2\times 10^{-12}$ &  $4.8\times 10^{-11}$ & $2.0\times 10^{-11}$ & 0.862 & 0.873 & 3.2$\times 10^{-12}$ &  $4.8\times 10^{-11}$ & $2.0\times 10^{-11}$ \\
5.0 &  0.94 &  0.92 & $4.5\times 10^{-9}$ &  $4.4\times 10^{-10}$ & $1.8\times 10^{-11}$ & 0.915 & 0.923 &   4.5$\times 10^{-9}$ & 4.4$\times 10^{-10}$ & 1.8$\times 10^{-11}$ \\
6.0 &  1.16 &  0.97 & $3.2\times 10^{-7}$ &  $8.3\times 10^{-10}$ & $5.3\times 10^{-11}$ & 0.973 & 0.978 &   2.3$\times 10^{-7}$ & 8.2$\times 10^{-10}$ & 5.3$\times 10^{-11}$ \\
7.0 &  1.23 &  1.05 & $5.6\times 10^{-6}$ &  $1.4\times 10^{-9}$ & $5.3\times 10^{-9}$ &  1.042 & 1.044 &   2.7$\times 10^{-6}$ & 2.0$\times 10^{-7}$ & 1.9$\times 10^{-9}$ \\
8.0 &  1.49 &  1.13 & $4.0\times 10^{-5}$ &  $1.6\times 10^{-7}$ & $3.9\times 10^{-5}$ &  1.134 & 1.136 &   $2.6\times 10^{-5}$ & $7.6\times 10^{-6}$ & 1.5$\times 10^{-7}$   \\ 
9.0 &  1.77 &  1.24 & $1.4\times 10^{-3}$ &  $3.8\times 10^{-5}$ & $3.4\times 10^{-4}$ &  1.240 & 1.241 &   9.1$\times 10^{-4}$ & 5.7$\times 10^{-4}$ & 3.4$\times 10^{-4}$   \\
\hline
\end{tabular}

\vspace{0.5cm}
$Z=10^{-5}$\\
\begin{tabular}{ccccccccccc}\\
\hline
   & Bef. SDU & Aft. SDU &  &  &  & 1$^{\rm st}$ TP &  &  &  & \\
\hline
${\rm {M_{ZAMS}}}$ & 
${\rm {M_{HexC}}}$ & 
${\rm {M_{HexC}}}$ &
$X_{\rm s}(C)$ & $X_{\rm s}(N)$ & $X_{\rm s}(O)$   &
${\rm {M_{HeexC}}}$ & 
${\rm {M_{HexC}}}$ &
$X_{\rm s}(C)$ & $X_{\rm s}(N)$ & $X_{\rm s}(O)$
\\
\hline
\hline
4.0 & 0.98  &  0.87 & $5.8\times 10^{-7}$ &  $3.1\times 10^{-6}$ & $3.9\times 10^{-6}$ &  0.862 &  0.875 &   $5.8\times 10^{-7}$ & $3.1\times 10^{-6}$ & $3.9\times 10^{-6}$   \\
5.0 & 1.16  &  0.91 & $5.3\times 10^{-7}$ &  $3.5\times 10^{-6}$ & $3.4\times 10^{-6}$ & 0.900  & 0.910 &   $5.3\times 10^{-7}$ & $3.6\times 10^{-6}$ & 3.4$\times 10^{-6}$   \\
6.0 & 1.52  &  0.97 & $4.9\times 10^{-7}$ &  $3.8\times 10^{-7}$ & $3.1\times 10^{-6}$ & 0.964   & 0.962  &   $5.2\times 10^{-7}$ & $3.8\times 10^{-6}$ & 3.2$\times 10^{-6}$   \\
7.0 &  1.69 &  1.05 & $4.1\times 10^{-6}$ &  $4.0\times 10^{-6}$ & $3.1\times 10^{-6}$ &  1.054 & 1.057 &   $1.2\times 10^{-6}$ & $7.4\times 10^{-6}$ & 3.0$\times 10^{-6}$   \\
8.0 &  1.96 &  1.18 & $2.7\times 10^{-5}$ &  $9.0\times 10^{-4}$ & $8.2\times 10^{-5}$ &  1.183 & 1.184 &   $2.7\times 10^{-5}$ & $9.0\times 10^{-4}$ & 8.2$\times 10^{-5}$   \\ 
9.0 &  2.25 &  1.33 & $1.6\times 10^{-3}$ &  $6.5\times 10^{-4}$ & $2.8\times 10^{-4}$ &  1.333 & 1.334 &   $8.6\times 10^{-4}$ & $1.2\times 10^{-3}$ & 4.0$\times 10^{-4}$   \\

\hline
\end{tabular}
\end{center}
\begin{center}
\caption{Relevant structure and composition parameters for the primordial and $Z=10^{-5}$ models. 
${\rm {M_{HexC}}}$ 
represents the mass of the H-exhausted core, and is given before and after 
the second dredge-up (SDU).  
$X_s(C)$, $X_s(N)$ and $X_s(O)$ in columns 4 to 6 are, respectively, the surface abundances of C, N and O after the SDU. 
${\rm {M_{HeexC}}}$ and ${\rm{M_{HexC}}}$ are, respectively, the masses of the He- and H-exhausted cores just before the first thermal pulse of the thermally-pulsing AGB or Super-AGB.
$X_s(C)$, $X_s(N)$ and $X_s(O)$ in columns 9 to 11 are, respectively, the surface abundances of C, N and O at this time.
}\label{tab:refe2}
\end{center}
\end{table*}

The results for the example models presented in this manuscript have been obtained with \textsc{monstar}, the Monash University Stellar Structure code (see for instance \citealt{frost1996}, \citealt{campbell2008}, \citealt{,gilpons2013}). It considers the isotopes relevant for the evolution ($^1$H, \iso{3}He, $^4$He, $^{12}$C, $^{14}$N, $^{16}$O and the rest of species are included in $Z_{\rm other}$). Nuclear reaction rates are from \cite{caughlan1988} with the update from NACRE \citep{angulo1999} for the ${\rm ^{14}N(p,\gamma)^{15}O}$. For discussion on implementation of carbon burning in a limited nuclear network we refer to \citet{doherty2010}.
The convective treatment implements the modified Schwarzschild criterion with the attempt to search for convective neutrality (\citealt{castellani1971}, \citealt{robertson1972}, \citealt{frost1996}), which is also known as induced overshooting.
This treatment intends to limit the effects of the unphysical discontinuity in the radiative gradient  at the convective  boundary that is induced by the composition difference between the mixed convective zone and the adjacent radiative shells (the details about this algorithm can be found in \citealt{frost1996}).

Mass-loss rates are calculated following \cite{vassiliadis1993}, and opacities are from OPAL \citep{igl96} for the interior, and either the molecular opacities from \cite{ferguson2005} for the $Z=10^{-10}$, $10^{-8}$ and $10^{-6}$ models, or \cite{lederer2009} and \cite{marigo2009} for the $Z=10^{-5}$ case.
Note that our models are solar-scaled, following \cite{gre93}, with $Z_{sun}=0.02$. 
Besides our primordial models use the initial metallicity from \cite{gilpons2005}, that is, $Z=10^{-10}$. This value is higher than the strict $Z=0$ frequently used in the literature (e.g. \citealt{chieffi2001}, \citealt{siess2002}), and the approximate values $Z=10^{-12}-10^{-13}$ expected from Big-Bang nucleosynthesis \citep{coc2014}. Nevertheless, as we will see later in this section, in terms of the characteristics of the evolution, yields and fates of the considered stars, $Z=0$ and $Z=10^{-10}$ produce the same results.
The limitations imposed by additional choices of input physics are discussed in due course.

Models have been computed for  $Z= 10^{-10}$ (primordial), $10^{-8}$ and $10^{-6}$, for initial masses between 3 and 9.8~\msun. Models for the $Z= 10^{-5}$ case with masses between 4 and 9~\msun{} were taken from \cite{gilpons2013}. An initial mass spacing of 1~\msun{} was chosen, except for cases near the mass thresholds for the formation of SNI1/2, where additional models were calculated  to obtain a mass spacing of 0.5~\msun, and for the cases near the mass thresholds for the formation of electron-capture and core collapse SNe, where we chose a mass spacing of 0.1~\msun.

\begin{figure}[t]
\begin{center}
\includegraphics[width=\columnwidth]{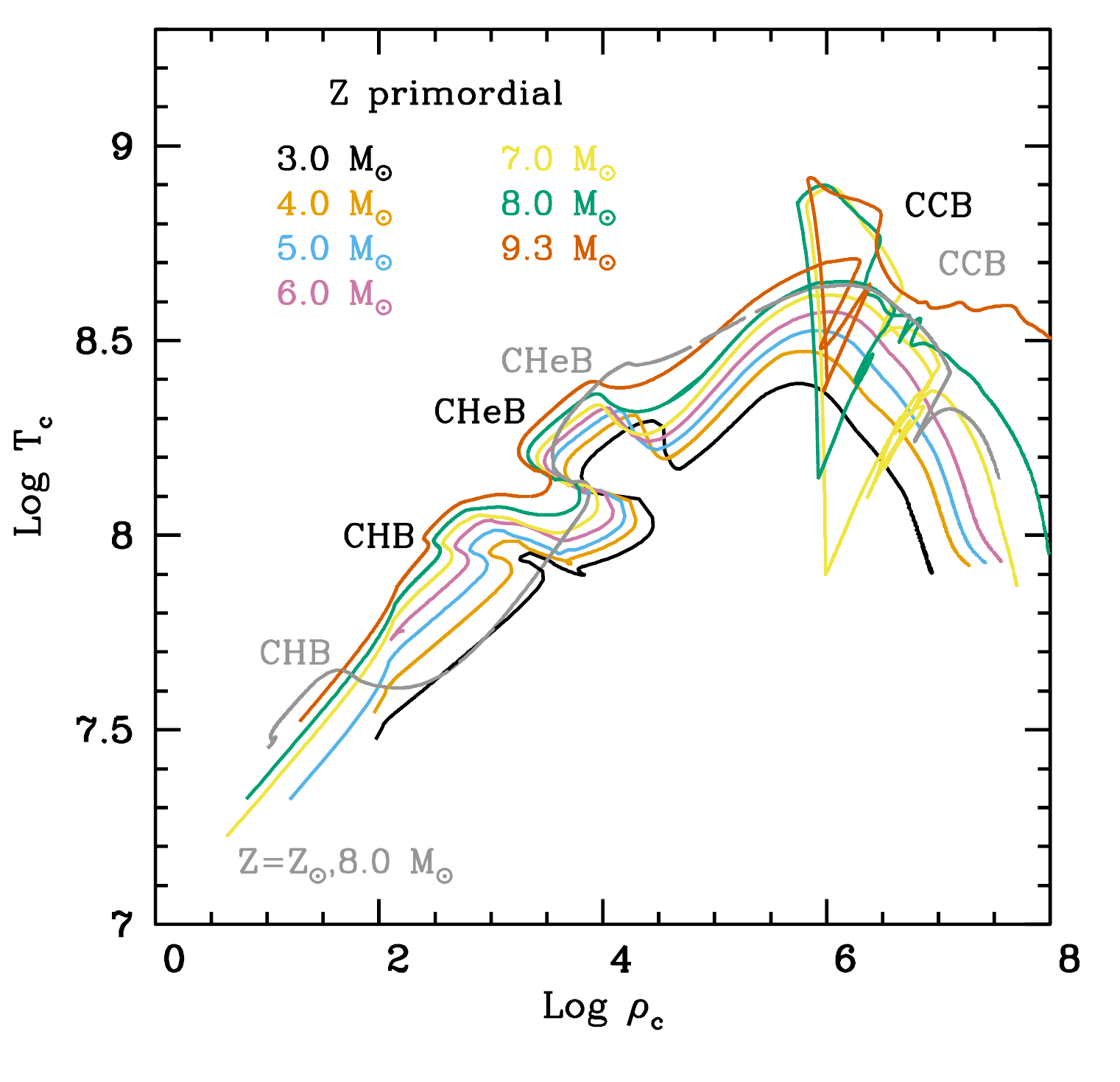}
\caption{Evolution in the $\log \rho_{\rm c}$-$\log T_{\rm c}$ plane of some selected models of primordial metallicity. The approximate locations of the main central burning stages H, He and C  are labelled CHB, CHeB and CCB respectively. For comparison we also show the evolution of the 8.0~\msun{} solar metallicity model (grey line and labels).} \label{fig:cenz0}. 
\end{center}
\end{figure}

\begin{figure}[t]
\begin{center}
\vspace{-.1cm}
\includegraphics[width=\columnwidth]{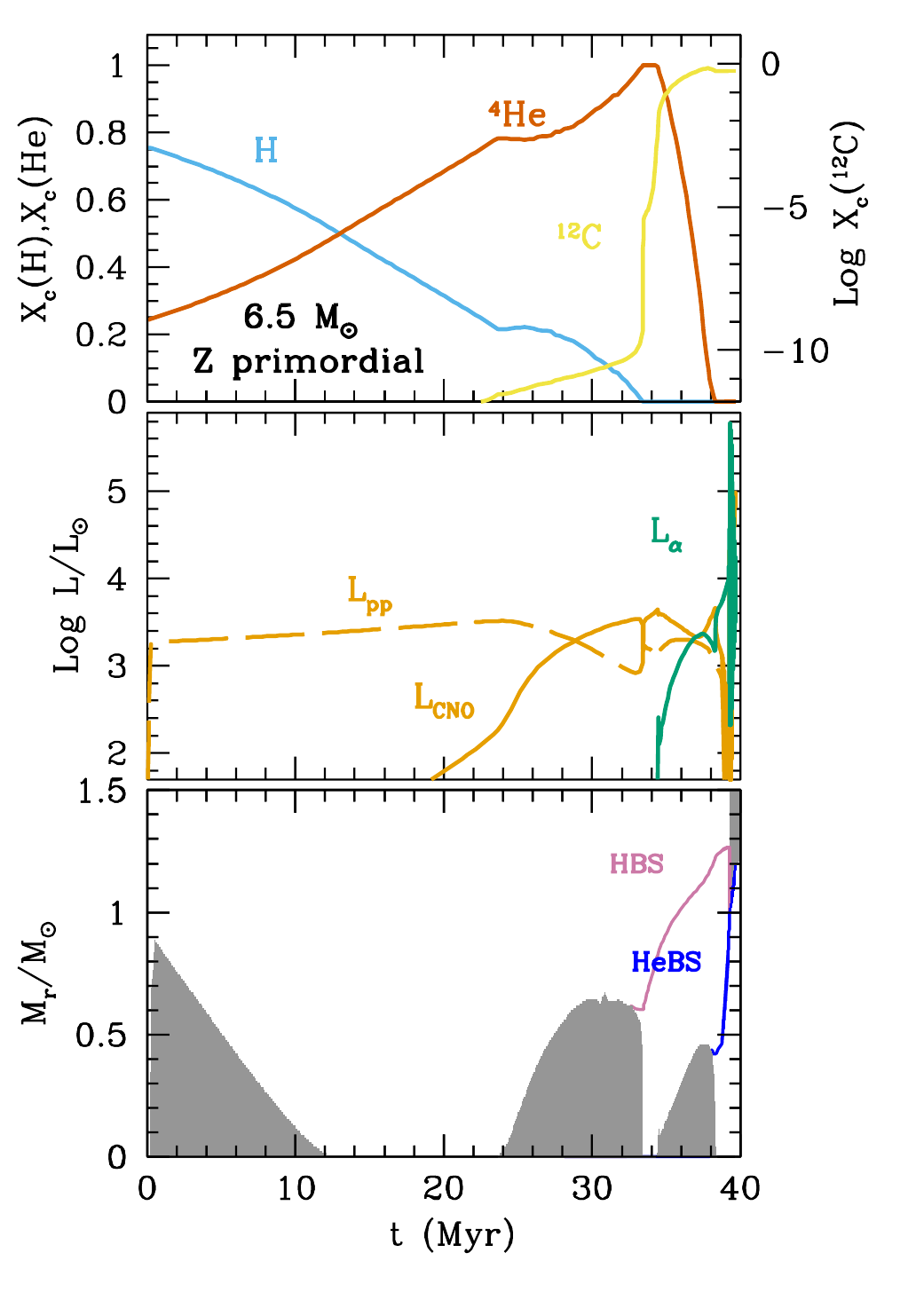}
\caption{Evolution of a 6.5~\msun{} primordial model. Upper panel: evolution of the central abundances of H,  $^4$He, and $^{12}$C . Middle panel: evolution of the luminosities from H-burning through the pp-chains (L$_{\rm pp}$), the CNO-cycle ($\rm L_{CNO}$), and the 3$\alpha$ reaction (L$_{\alpha}$). Lower panel: evolution of convective zones and the location of the H-burning shell (HBS) and of the He-burning shell (HeBS). }
\label{fig:corez0}
\end{center}
\end{figure}

\begin{figure}[t]
\begin{center}
\vspace{-0.1cm}
\includegraphics[width=\columnwidth]{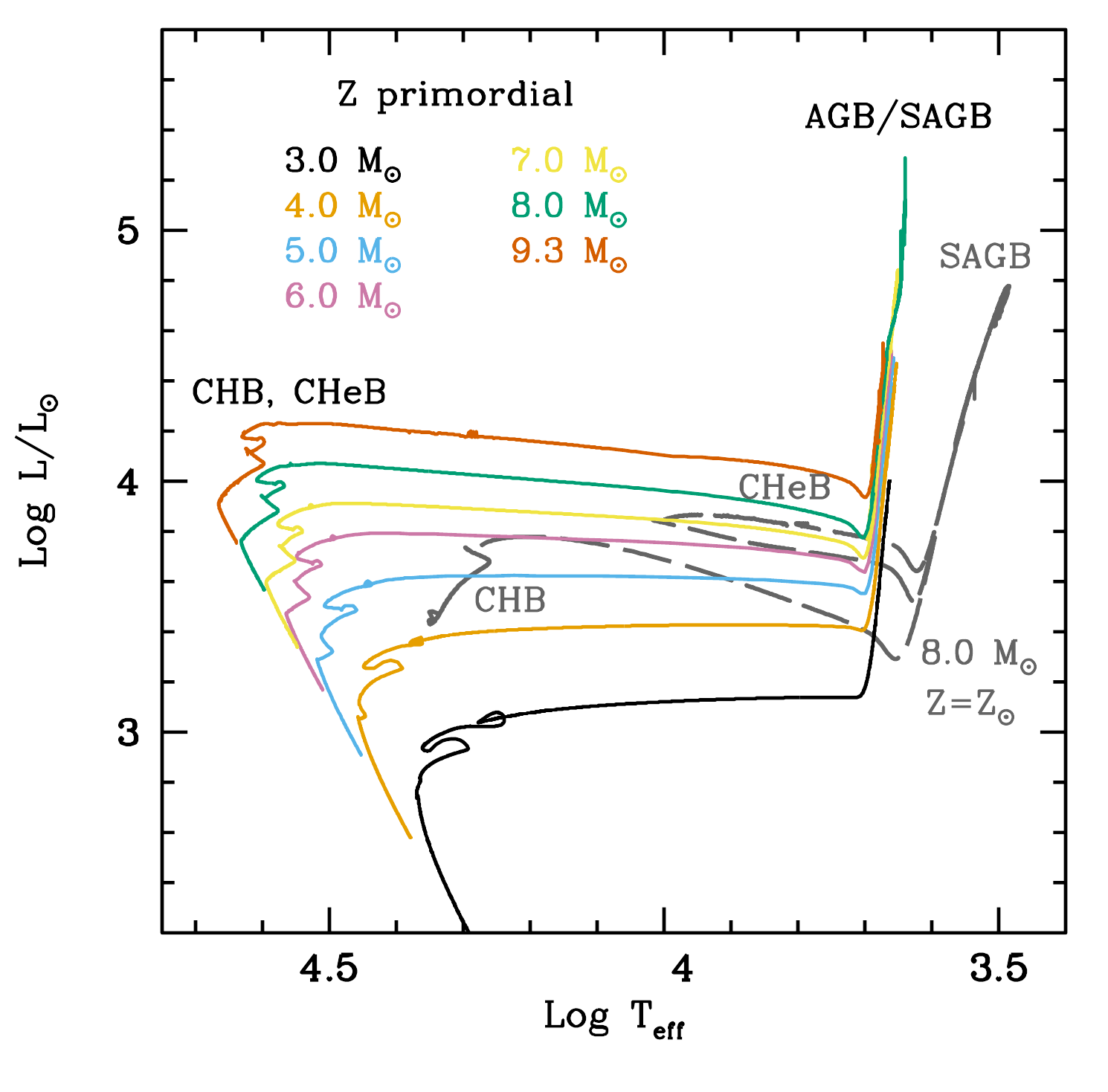}
\caption{Evolution in the Hertzsprung-Russell diagram of some selected models of primordial metallicity.
The approximate locations of the main central burning stages are labelled.
For comparison, the evolution of an 8.0~\msun{} solar metallicity model has been included. The evolution along the thermally-pulsing AGB or Super-AGB has been truncated for better display.}
\label{Fig_hrz0}
\end{center}
\end{figure}

\subsection{Evolution during the main central-burning stages}

\subsubsection{Core hydrogen and helium burning}
Stars that will become Super-AGB stars are at the upper end of the mass range defined as intermediate mass stars (IMS). We will refer to these stars, destined to become Super-AGB stars, as SIMS for Super Intermediate Mass Stars. We save the name Super-AGB for that specific phase of evolution of the SIMS.
The evolution of primordial and EMP IMS presents substantial differences with respect to that of higher $Z$ objects.  
The main central burning stages of primordial stars over a wide range of masses have been well known since the 1970s (see section \ref{sec:intro} for references).  
The absence of metals and, in particular, of C and N forces the star to ignite central H through the pp-chains and form a relatively small convective core. Because the energy generation rates associated with the pp-chains ($\propto T^n$ with $n\simeq 4$) are more weakly dependent on temperature than those associated with the CN-cycle (with $n\simeq 20$), main sequence primordial stars are more compact and hotter than their higher $Z$ counterparts of similar masses (see Figure \ref{fig:cenz0}). 
Central H-burning temperatures in primordial models reach values $\sim$ 10$^8\:$K, whereas those of solar metallicity remain $\lesssim$ 4$\times$10$^7\:$K.
During core H burning both the central temperature and density smoothly increase and allow the synthesis of He and a small amount of C, via the triple-alpha reaction. Note at this point that the strong temperature dependence of the 3-$\alpha$ reaction rate (roughly $\propto T^{40}$), together with the high central temperatures during core H-burning, are critical to understanding the formation of $^{12}$C in these primordial stars. Once the total mass fraction of C reaches $\sim 10^{-10}$ the CN-cycle starts operating, which causes a sudden increase in the release of energy, a brief core expansion period and the disappearance of core convection. After the core readjusts itself, central H-burning continues and is now dominated by the CN-cycle. The central density and temperature rise again and a new convective core forms and lasts until the end of core H-burning. The particular value of the central C abundance at the onset of the CN-cycle, the duration of the entire core H-burning phase, and the resulting mass of the H-exhausted core strongly depend on the adopted input physics, such as the nuclear reaction rates, the assumptions concerning convective overshooting and the choice of opacity tables \citep{siess2002}.
In general, all models of initial mass above 1~\msun{} experience the transition from pp-chain to CN-cycle dominated core H-burning. This transition occurs earlier (and thus with higher central H-abundance) for more massive models.

As an example of central H and He-burning stages, we show the evolution of a primordial 6.5~\msun{} model in Figure \ref{fig:corez0}.

The evolution of central temperature versus central density (${\log \rho_{\rm c}-\log T_{\rm c}}$) for some selected models of primordial intermediate-mass stars and, for comparison, the evolution of an 8.0~\msun{} solar metallicity case are shown in Figure \ref{fig:cenz0}. In this figure the occurrence of core H-burning at higher T for the primordial cases can be clearly seen. Once central H is exhausted, the structure and composition of the resulting He cores are similar to analogous cores from higher $Z$ stars and thus both the core He and C burning phases 
occur at similar loci in the $\log \rho_{\rm c}-\log T_{\rm c}$ diagram.
Indeed, even if the physical evolution of the He-core does not directly depend on 
its metallicity, it is indirectly influenced through the behaviour of the H-burning shell.
Intermediate-mass H-exhausted cores are more compact and hotter than their higher $Z$ counterparts. Therefore, central He-burning starts and the central 3$\alpha$ reactions provide energy supply very shortly after core H-burning (\citealt{chieffi2001}, \citealt{siess2002}). Consequently stellar contraction stops, the star stays in the blue region of the HR-diagram, and an efficient H-burning shell does not develop. 
Without a powerful H-burning shell the corresponding envelope expansion and cooling associated with the ascent of the RGB are avoided. The high temperature gradients which would drive the formation of a deep convective envelope are not achieved and thus the first dredge-up process is also averted\footnote{The actual occurrence or avoidance of the RGB is actually quite a complex phenomenon and depends on many factors (e.g. \citealt{sugimoto2000}, \citealt{stancliffe2009}).}. Thus intermediate-mass primordial stars maintain a pristine envelope until the end of core He-burning.\footnote{Note that low-mass primordial models  ($\rm M_{ZAMS}\lesssim$ 1.3~\msun{}) show a different behaviour. They climb the RGB and ignite He off-centre in conditions of partial degeneracy. As a consequence they develop a He-flash,  followed by a H-flash and a proton-ingestion episode (e.g. \citealt{fujimoto2000}, \citealt{schlattl2001}, \citealt{picardi2004}).}

Table \ref{times} shows the approximate lifetimes (at the end of calculations) of a selection of EMP model stars. We clearly see the reduction of stellar lifetimes with decreasing metallicity. 
The differences between these lifetimes and those given by \cite{siess2002} are small, being between 0.4\% and 4\%.

\begin{figure*}[t]
\begin{center}
\includegraphics[width=0.85\linewidth]{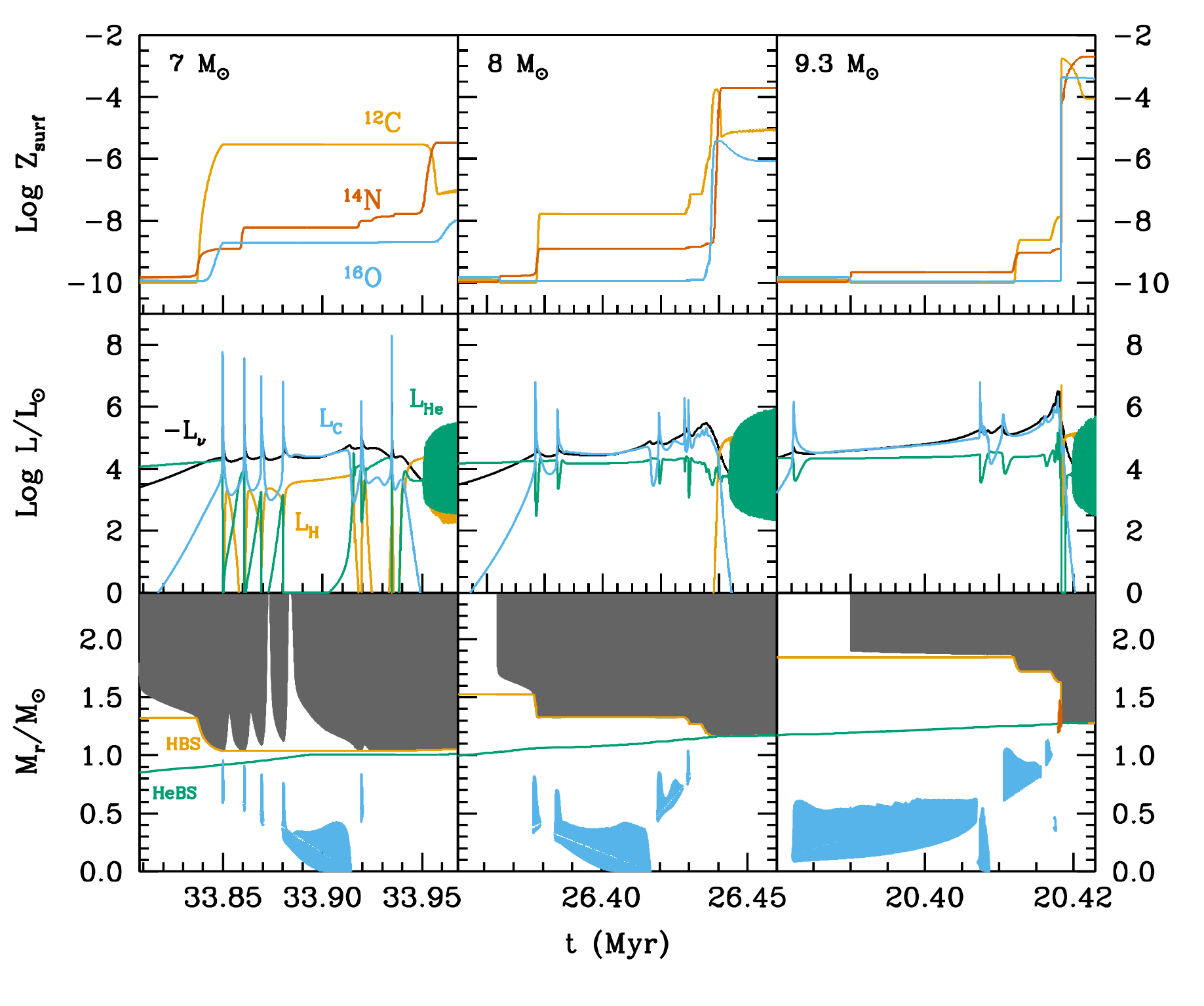}
\caption{Summary of the evolution during C burning (starting near the beginning of the Early-AGB phase), and the first thermal pulses of the thermally-pulsing Super-AGB for the 7, 8 and 9.3~\msun{} models with primordial $Z$. Lower panels show the temporal evolution of the convective envelope (grey) and of the inner convective shells (the ones associated to C flashes are shown in blue, and the one associated to He-burning and gravothermal energy release during the dredge-out episode of the 9.3~\msun{} model is shown in vermilion). 
We also show the evolution of the mass location of the H-burning shell (orange) and the He-burning shell (green). 
Middle panels show 
the evolution of the luminosities from H, He, and C burning together with neutrino losses (${\rm L_H}$, ${\rm L_{He}}$, ${\rm L_C}$ and  ${\rm L_{\nu}}$ respectively). Upper panels show the evolution of surface mass fractions (Z$_\mathrm {surf}$) of C, N and O.}
\label{fig:panelz0}
\end{center}
\end{figure*}

The avoidance of the first dredge-up is not a phenomenon unique to intermediate-mass primordial stars, as it is also 
shared by intermediate-mass stars of initial metallicity lower than $Z_\mathrm{ZAMS} \sim 10^{-3}$.
The evolution in the Hertzsprung-Russell diagram of some models of primordial IMS and, for comparison, also a solar metallicity IMS, are shown in Figure \ref{Fig_hrz0}. 
Both the core H-burning and the core He-burning phases occur in the hot part of this diagram for the primordial metallicity objects. They also evolve at higher luminosities until the AGB or Super-AGB phase, and remain hotter during this phase \citep{becker1977}. 
At this point a new overall contraction ensues, an efficient H-burning shell finally forms, and the star expands and cools to become a giant hosting a deep convective envelope. Then the second dredge-up process begins. Note that intermediate-mass primordial stars do not develop a first dredge-up, but the terminology of second dredge-up is still used in the literature to refer to the dredge-up episode occurring at the end of core He-burning, by analogy with higher $Z$ cases. We will show in section \ref{sec:sdu} that the 
efficiency of this process is very sensitive 
to the choice of input physics (and associated uncertainties). This is
critical for the later evolution as thermally-pulsing AGB or Super-AGB stars and, eventually, for their final fates.
Tables \ref{tab:refe1} and \ref{tab:refe2} show a summary of relevant parameters during the evolution of a selection of our primordial and $Z=10^{-5}$ models.

\subsubsection{Carbon burning} \label{cburning}
Regardless of their initial metallicity, all stars that develop CO cores of masses $\gtrsim$ 1.05~\msun{} after central H- and He-burning will proceed to the ignition of carbon.  
It is important to recall that the central C abundance at the time of ignition critically depends on the characteristics of the previous He-burning phase 
and, in particular, on the occurrence of breathing pulses 
\citep{castellani1985}, a type of convective instability which occurs near the 
time of central He-exhaustion, and affects the convective core boundary. 
Their extent, and even their occurrence, strongly depend on the the numerical 
treatment of convective boundaries \citep{constantino2017}.

Carbon-burning in primordial to $Z=10^{-5}$ stars occurs in a very similar fashion to their higher metallicity counterparts (\citealt{gilpons2005}, \citealt{gilpons2013}). The details of the process have been known since the 1990s
\citep[][and references therein]{ritossa1999}, 
with ignition occurring in conditions of partial degeneracy for solar metallicity in intermediate-mass stars. This was further analysed in, e.g. \cite{siess2006}, \cite{doherty2010}, \cite{farmer2015} and references therein. Here we present a brief overview, highlighting the few particularities of metal-poor stars, and refer to \cite{doherty2017} for more detail.

\begin{figure*}[t]
\begin{center}
\includegraphics[width=0.8\linewidth]{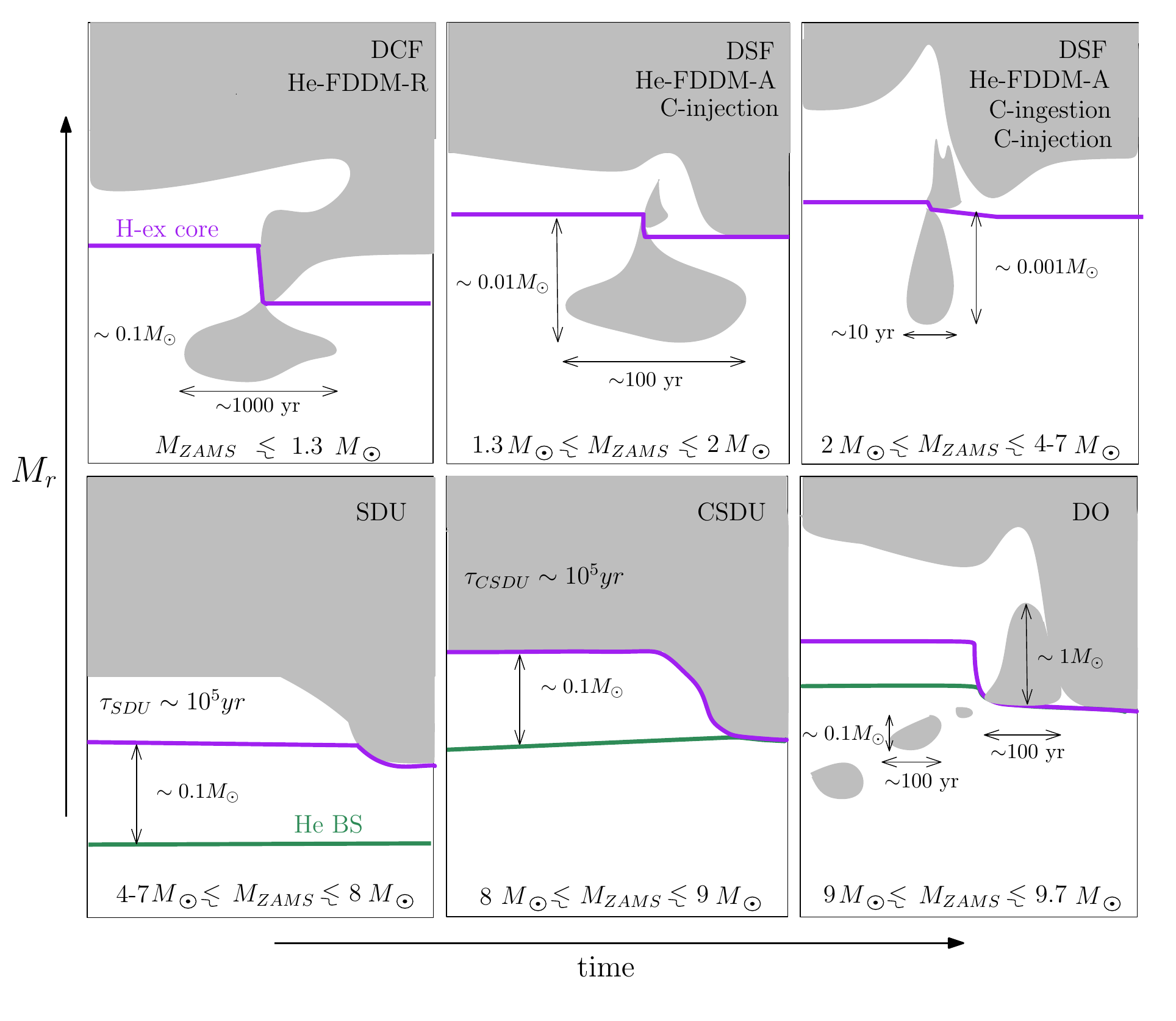}
\caption{Schematic view of mixing episodes in metal-poor stars. The grey areas show the location of convective zones in the mass coordinate ${\rm M_r}$ versus time, the purple line shows the outer limit of the H-exhausted core, 
(defined as the mass coordinate where the H mass fraction drops below $10^{-6}$)
and the green line shows the location of the He-burning shell. Upper panels show the different nomenclature used to refer to the mixing phenomena. The upper left panel shows the dual core flash \citep{schlattl2001,picardi2004,campbell2008} or He-flash driven deep mixing event at the tip of the RGB \citep{suda2010}. The upper middle panel shows the dual shell flash \citep{campbell2008} or He-flash driven deep mixing event at the AGB \citep{suda2010}, also named C injection by \cite{siess2002}.
The upper right panel shows the He-flash driven deep mixing event at the AGB \citep{suda2010}, or proton ingestion (\citealt{chieffi2001}, \citealt{lau2008}, \citealt{cristallo2009} and \citealt{siess2002}). 
The lower left panel shows a standard second dredge-up episode (SDU), the lower middle panel shows a corrosive second dredge-up episode (CSDU), and the lower right panel shows a dredge-out episode, DO \citep{gilpons2013}. The orders of magnitude of the duration of the convective shell episodes and their sizes are given, as well as the orders of magnitude of the duration of the entire SDU and CSDU.}
\label{fig:mixing}
\end{center}
\end{figure*}

Figure \ref{fig:panelz0} summarises the evolution of the main structural parameters and the surface abundances of C, N and O for the 7, 8 and 9.3~\msun{} primordial models during C burning and the first thermal pulses of the 
Super-AGB phase. The models shown are, respectively, representative of low mass Super-AGBs, intermediate-mass Super-AGBs, and massive Super-AGB stars. 
Extended C burning occurs in stars which are able to form CO cores of masses $\gtrsim 1.05$~\msun{} and proceeds as follows. Once the central He-burning phase has been completed, the resulting CO core contracts and heats, so that neutrino energy losses become important for the innermost regions of the star. 
The temperature maximum  moves outward and when it reaches $\approx 6\times 10^8$K, carbon ignites off-centre (the higher the initial mass of the SIMS, the closer to the centre is the ignition). 
Because C burning takes place under conditions of partial degeneracy we find that the thermal instability produces strong flashes with peak luminosities that may exceed $10^8\,\mathrm{L}_{\odot}$, as seen in the middle panels of Figure~\ref{fig:panelz0}. 
Each flash provides large energy injections able to drive the formation of local convective zones which disappear shortly after the flashes are extinguished (see lower panels of Figure~\ref{fig:panelz0}). Successive flashes advance towards deeper regions of the core and, eventually, the C burning flame reaches the centre. Yet, the central temperature is not high enough for complete exhaustion of central C. The exceptions are the most massive SIMS, which burn C in an approximately stationary way and do exhaust central carbon completely, or leave a residual C abundance not higher than a few tenths of a percent.
C burning in Super-AGB stars is therefore similar to He-burning through core flashes in low mass-stars.
However, because in Super-AGB stars the CO core is more massive and the conditions there are more extreme, C burning must consume a larger amount of fuel than He-burning in low-mass stars to lift the degeneracy.
The C burning process does not finish when the C flame reaches the centre of the star. Instead, the CO degenerate regions located above the resulting ONe core also ignite in flashes and develop associated convective shells. 
At the end of C burning a typical early Super-AGB star is composed of an ONe rich core, a CO-rich shell and a H and He-rich envelope, more or less polluted in metals by the effect of the different mixing episodes that
we will describe in the next subsection. 

The location of the base of the convective envelope is altered during C burning because of the highly energetic C flashes.
These flashes drive local expansion and cooling which causes the recession of the 
convective envelope. Once the thermal conditions that existed prior  to the flashes are restored, the bottom of the convective envelope returns close to its position before the occurrence of the flash.

The minimum mass for C ignition, referred to as $\rm M_{up}$ depends on the composition, input physics and on numerical aspects of the evolutionary calculations. With the physical prescriptions adopted here, \textsc{monstar} yields a lower mass threshold of 6.8~\msun{} for the primordial star, and the corresponding model experiences five convective flashes before C burning reaches the centre. At the time of carbon ignition the partially degenerate CO-core mass is 
1.05~\msun, and the central carbon abundance is 0.55. C ignition is located at the mass point 0.69~\msun.
We are following the definition of $\rm M_{up}$ proposed by \cite{doherty2015}, which requires the formation of a C convective shell.
As a comparison, the 6.7~\msun{} model experiences C burning briefly and ineffectively, with associated maximum luminosities of only a few hundred L$_{\odot}$, without C convective shells, and resulting in a practically unaltered CO core. 

The highest mass for which a primordial star experiences the Super-AGB phase is $\sim$ 9.7~\msun. This model ignites C very close to the centre, in conditions of degeneracy much milder than those of the 6.8~\msun{} model. 
Note that the lowest initial mass for the occurrence of central C ignition does not correspond to the upper mass threshold for the occurrence of Super-AGB stars. Instead, some stars may ignite C centrally, develop a brief inefficient Ne-burning phase, and continue their lives as thermally-pulsing  Super-AGB stars.

\subsection{Mixing episodes prior to the thermally-pulsing AGB or Super-AGB phase}\label{sec:prevmixing}

Prior to the thermal pulsing phase, a variety of mixing processes enrich the stellar surface in metals.
The present work focuses on intermediate-mass evolution and thus, in the following subsections, we describe the second dredge-up and the dredge-out episodes. However, it is also appropriate to mention the occurrence of a proton-ingestion episode (PIE) during the core He-flash, located at the tip of the red giant branch (RGB) for low-mass stars ($\rm M_{ZAMS} \lesssim 1.3$~\msun{}). 
PIEs result from rapid ingestion of protons into high temperature regions, typically regions where He-burning is active. 
Through their modelling of a low-mass primordial star \cite{dantona1982} originally speculated that these types of events may occur. This was confirmed by \cite{fuj90} and \cite{hollowell1990}, and has been studied regularly since then (e.g.  \citealt{cassisi1996}, \citealt{fujimoto2000}, \citealt{schlattl2001}, \citealt{picardi2004}, \citealt{campbell2008}, \citealt{mocak2010}, \citealt{suda2010}, \citealt{lugaro2012}, \citealt{cruz13} and references therein). 
 Even though they share common features, the Dual Core Flashes (DCF) that occur at the tip of the RGB are different from the Dual Shell Flashes (DSF) that develop in intermediate-mass stars at later times during the thermally-pulsing AGB, and involve He-convective zones associated with thermal pulses (see Section \ref{sec:tpagb}). For the sake of clarity, the relevant mass-ranges and the different nomenclature for various mixing events are shown in Figure \ref{fig:mixing}.
 
\begin{figure}[t]
 \begin{center}
 \vspace{-0.4cm}
 \includegraphics[width=8.8cm,angle=0]{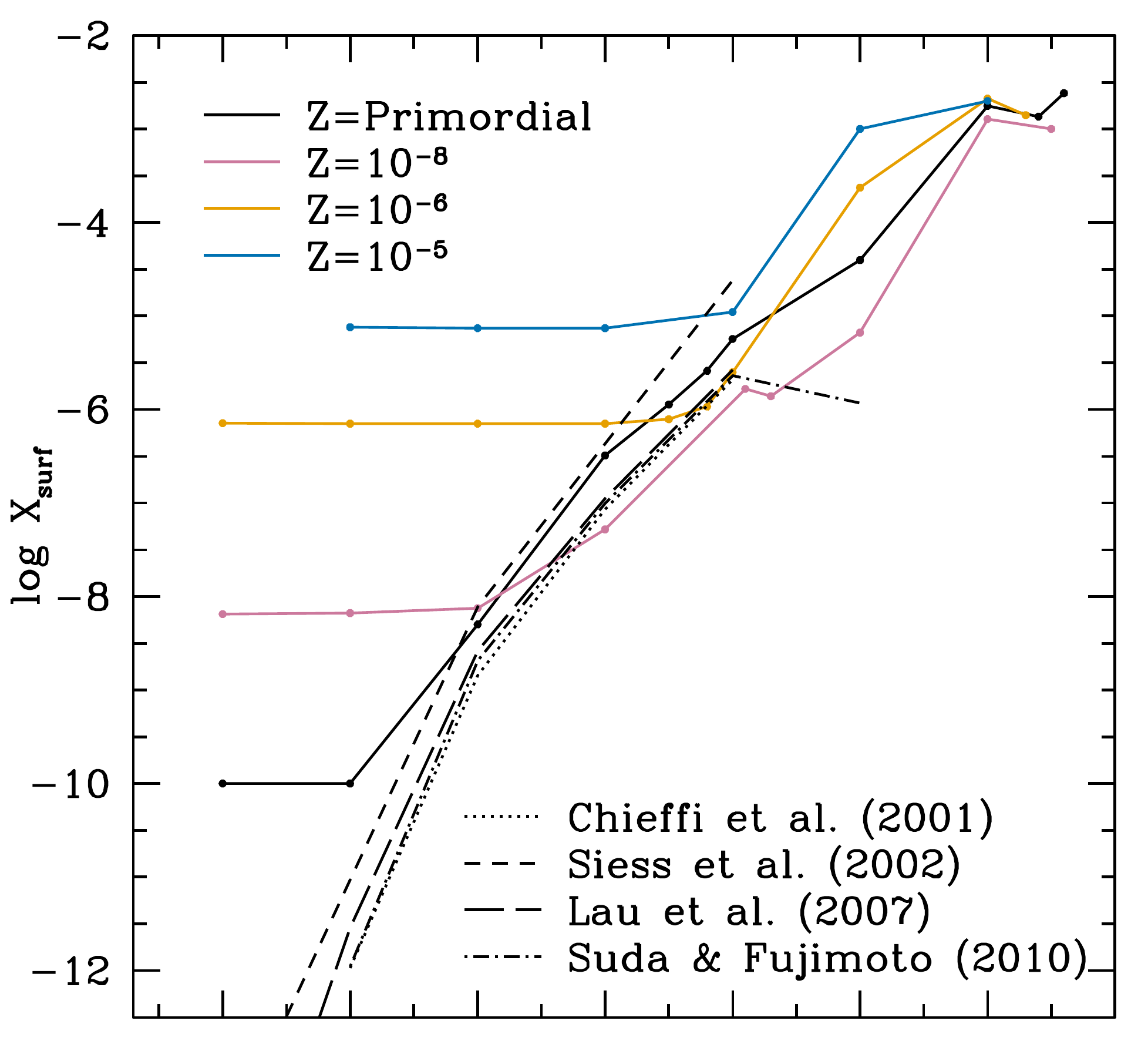}
 \includegraphics[width=8.8cm,angle=0]{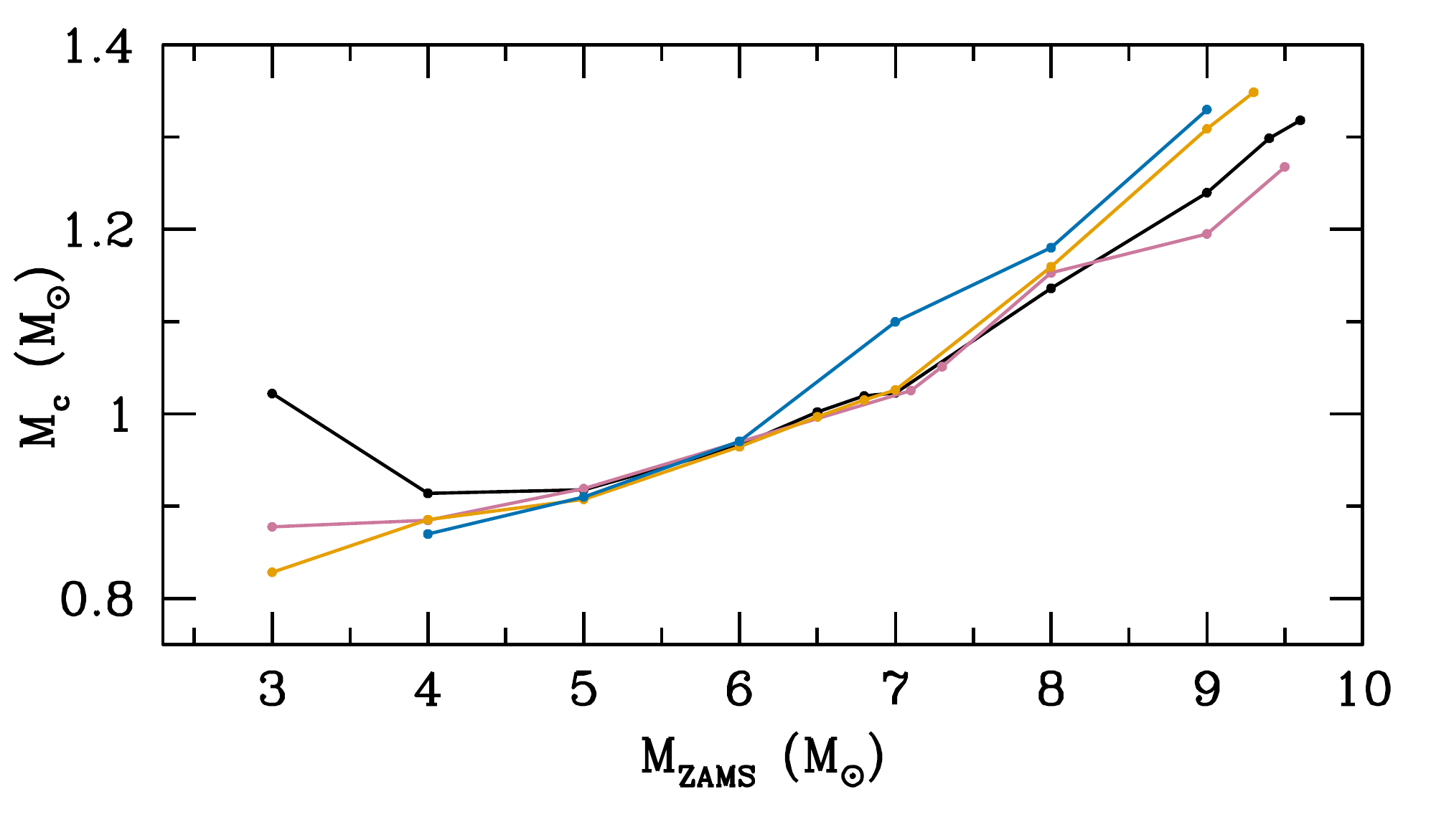}
 \caption{Upper panel: second dredge-up episode enrichments for primordial to $Z=10^{-5}$ model stars. Solid lines correspond to models computed with \textsc{monstar}. $X_\mathrm{surf}$ represents the sum of the mass fraction of all species with 
 atomic number $\geq$ 6. Note that primordial models in this case have been computed with $Z_\mathrm{ZAMS}=10^{-10}$ (see text for details). The primordial models by other authors use $Z_\mathrm{ZAMS}=0$. Bottom panel: size of the H-exhausted core ${\rm M_c}$ at the end of the second dredge-up. }
 \label{fig:sdu}
 \end{center}
 \end{figure}

\subsubsection{The Second Dredge-Up}\label{sec:sdu}

For stars of initial metallicity $Z \lesssim 10^{-3}$ the first ascent of the giant branch occurs after the exhaustion of central He. In a normal second dredge-up episode the envelope expansion is accompanied by the formation of a deep convective envelope, able to penetrate the He-core. This second dredge-up episode results in envelope enrichment of $^4$He, $^{14}$N and $^{13}$C, and depletion in $^{12}$C and, to a lesser extent, $^{16}$O. In the case of primordial to $Z=10^{-8}$ stars,  many models experience primarily an increase in the $^{12}$C and $^{16}$O surface abundances 
(see \citealt{lau2009} and references therein). Although the changes to the surface composition are similar, they are driven by different processes. 

For the lowest metallicities there is a relatively low entropy barrier and a higher compactness and temperature. In particular, the high temperatures in the H burning shell (near $10^8\:$K) allow the occurrence of the 3$\alpha$ reaction within this shell (see, for instance, \citealt{chieffi2001}). When this material is engulfed by convection and dredged to the surface it results in increases in the abundance of $^{12}$C and $^{16}$O. 
Even though the result in terms of surface composition is similar (an increase in $^{12}$C and $^{16}$O), we should distinguish this type of {\it hot second dredge-up episode} from the {\it corrosive} second-dredge-up reported for the more massive $Z=10^{-5}$ stars in \cite{gilpons2013} (see Figure \ref{fig:mixing}). In the corrosive second dredge-up the base of the convective envelope is able to dredge-up material from the CO core. The corrosive second dredge-up is actually present for initial masses $\gtrsim$ 8~\msun{} in the metallicity range from primordial to $Z=10^{-4}$, but also up to solar metallicity in narrower mass ranges \citep{doherty2015}. 
Note that during this event the He-burning shell remains active, with a He luminosity of a few thousands \lsun.

\cite{lau2009} presented detailed post-second dredge-up surface abundances of intermediate-mass stars (2--6~\msun) of metallicities between $Z=10^{-8}$ and $Z=10^{-4}$.
They found a very mild enrichment in their $10^{-8} \le Z \le 10^{-7}$ models for $\rm M_{ZAMS}\lesssim 5$~\msun{} but a significant pollution (up to $Z_\mathrm{surf} \sim 10^{-6}$) for their 6~\msun{} model. 
In the metallicity range $10^{-6}\leq Z \le 10^{-4}$ the largest surface enhancement occurred for models with $\rm 3~M_\odot \leq M_{ZAMS} \leq 5~M_\odot$.
This metal pollution is due to the hot second dredge-up described above. Additionally, \cite{lau2009} showed that the implementation of overshooting below the envelope (treated as in \citealt{schroder1997}, with $\delta_{ov}=0.12$), further increased second dredge-up efficiency, and they calculated the corresponding surface abundances.

A summary of surface abundances after second dredge-up obtained by different authors is presented in Figure \ref{fig:sdu}. We also present the resulting core masses and surface metal abundances obtained with \textsc{monstar}, after the second dredge-up, corrosive second dredge-up, or dredge-out (explained in the next subsection). Note that the primordial 3~\msun{} model does not undergo a second dredge-up episode.
Note also that the precise initial metallicity for the primordial cases in our example models ($Z=10^{-10}$) is different in the models from the literature ($Z=0$ strictly).

The details of the treatment of convective boundaries and mixing are particularly critical for the second dredge-up and the later evolution and fate of primordial to EMP SIMS of $\rm M_{ZAMS} \gtrsim$ 7--9~\msun. Stellar models which implement the strict Schwarzschild criterion undergo a rather mild second dredge-up \citep{suda2010}, whereas the inclusion of overshooting produces a higher surface enrichment  (see \citealt{chieffi2001}, \citealt{siess2002}). The calculations with \textsc{monstar} presented in this work, which implement a treatment of convection that includes the search for neutrality (\citealt{JL86} and \citealt{frost1996}), also lead to a moderately high enrichment in surface metals.

In the case of Super-AGB stars, second dredge-up occurs at different stages of the C burning phase for stars of different initial masses. 
For primordial to EMP stars up to $\approx$ 7~\msun{} (destined to become low-mass Super-AGB stars) it takes place before the first C flash, and its effects are relatively mild. As an example, the primordial 7~\msun{} star envelope is enriched only up to a metallicity of $Z_\mathrm{surf} \sim 10^{-6}$. Stars of higher initial mass have hotter He-exhausted cores and thus ignite C earlier. For instance, the 8~\msun{} model experiences the corrosive second dredge-up after the first C flash. This is shown in the upper middle panel of Figure \ref{fig:panelz0}, in which the C surface abundance of the 
8~\msun{} model peaks to values above $10^{-4}$ shortly before the thermally-pulsing Super-AGB begins.
Finally, the envelopes of the most massive Super-AGB stars, such as the primordial 9.3~\msun{} in Figure \ref{fig:panelz0}, are only enriched at the end of the C burning process, and shortly before the dredge-out occurs.

\subsubsection{Dredge-out episodes}\label{sec:dout}

The most massive Super-AGB stars ($\gtrsim$ 9.2~\msun{} for the primordial case and $\gtrsim$ 8.8~\msun{} for the $Z=10^{-5}$ case) experience a type of proton-ingestion episode at the end of their C burning phase, in which a convective He-burning shell merges with the convective envelope. This so-called dredge-out process has been widely studied (\citealt{iben1997}, \citealt{ritossa1999}, \citealt{siess2007}, \citealt{gilpons2013}, \citealt{takahashi2013}, \citealt{doherty2015}, and \citealt{jones2016a}). 
During the dredge-out protons are ingested in regions of temperatures $\gtrsim 10^8$ K, in which He-burning is active, and thus a strong H-flash develops.
An example of a dredge-out episode is shown in the right panels of Figure \ref{fig:panelz0}. The behaviour of convective zones during this process is also outlined in Figure \ref{fig:mixing}.
The H-flashes associated with these proton-ingestion episodes are stronger for the highest initial mass cases (up to $10^{10}\: L_{\odot}$ for the primordial 9.5~\msun{} model). From a nucleosynthetic point of view, they are able to dredge-out very significant amounts of C and O to the stellar surface, whose metallicity increases from practically negligible to values above $Z=10^{-3}$. It is also worth noticing (see Figure \ref{fig:sdu}) that the final surface metallicity $Z_\mathrm{surf}$ after the dredge-out is practically the same for all metal-poor models, regardless of the initial $Z$.

Although dredge-out has been recognised since the 1990s, 
its effects on the star, and especially the nucleosynthesis, are far from well understood. This is primarily because the timescale for the ingestion of protons is similar to that of the burning of the very same protons. \cite{jones2016a}
suggested that the vast amount of energy that is
generated in a very narrow region during the H-flash might lead to an important mass ejection. i.e. the event may become hydrodynamical. This interesting hypothesis should be checked by 3D hydrodynamical calculations.

\subsection{Evolution during the thermally-pulsing AGB and Super-AGB phase}\label{sec:tpagb}

Once the main central burning stages are completed, intermediate-mass stars become giants consisting of a degenerate core (composed either of CO, CO-Ne, or ONe with a surrounding thin CO-shell), and a H-rich convective envelope. In either case both the H burning shell and the He burning shell become active and, as the He burning shell advances outwards and gets close enough to the H burning shell, a He-flash or thermal pulse ensues. This marks the beginning of the thermally-pulsing AGB or Super-AGB phase, in which steady H-burning and unstable He-burning alternate to provide the nuclear energy supply for the star. The thermally-pulsing AGB phase was recently described in detail in \cite{karakas2014}, and in \cite{doherty2017}, who placed special emphasis on the evolution of thermally-pulsing Super-AGB stars. Besides their characteristic double-shell burning, thermally-pulsing AGB and Super-AGB stars present additional features, such as the formation of inner convective shells, which are a consequence of the high and fast energy release occurring during each thermal pulse. From a nucleosynthetic point of view, thermally-pulsing 
AGB and Super-AGB stars may experience the phenomena known as hot-bottom burning and third dredge-up.

Primordial to EMP models of initial mass $\rm M_{ZAMS} \gtrsim 2-3$~\msun{} may experience hot bottom burning (\citealt{siess2002}, \citealt{lau2009}, \citealt{constantino2014}). One should note, however, that the occurrence of hot bottom burning as a function of  initial mass in the primordial to $Z=10^{-8}$ cases shows a peculiar behaviour, which will be 
analysed in the following subsections. 
Hot bottom burning is characterised by very high temperatures at the base of the convective envelope, especially in metal poor stars that develop more massive cores than their metal rich counterparts. The temperatures can reach extreme values $\gtrsim 160 \times 10^6$\,K and strongly impact the envelope composition (see Section \ref{sec:nucleosynthesis}).

The third dredge-up may occur after a thermal pulse and corresponds to the penetration of the convective envelope into the intershell region that contains material previously processed by He-burning. This third dredge-up causes surface enrichments in 3$\alpha$ products and has a direct impact on the fate of stars, as it alters the core-growth rate by repeatedly reducing the mass of the H-exhausted core. The third dredge-up may actually stop the stellar core from reaching the Chandrasekhar mass during the thermally-pulsing AGB or Super-AGB phase. 
Additionally, in EMP stars, the C surface enhancement caused by the third dredge-up may result in a significant increase in mass-loss rates. Unfortunately, the efficiency of this process 
and even its occurrence, is a matter of debate. Authors who computed and analysed the thermally-pulsing AGB and Super-AGB of primordial stars of masses  $\rm M_{ZAMS} \gtrsim$ 5~\msun{} either found quite efficient third dredge-up when using some degree of overshooting \citep{chieffi2001,siess2002}, or no third dredge-up at all when using the strict Schwarzschild criterion to determine the limits of convection (see for instance, \citealt{gilpons2007}, \citealt{lau2008}, \citealt{suda2010}), or even when applying some amount of overshooting  \citep{gilpons2007}.

\subsubsection{Do primordial and EMP AGB and Super-AGB stars experience thermal pulses?}

\cite{chieffi1984} were the first to perform
calculations beyond the main central burning stages of intermediate-mass primordial stars. They considered a 5~\msun{} model which developed a 0.78~\msun{} degenerate core. Unlike similar models of higher initial metallicities, their primordial star did not develop  He-flashes characteristic of the thermally-pulsing AGB phase.

Instead they found that He-burning proceeds steadily, and this behaviour was understood as a consequence of the higher temperatures of the H burning shell. In the absence of CNO elements, H is burnt at much higher temperatures, allowing for simultaneous production of carbon via the $3\alpha$ reactions. i.e. the 3$\alpha$ reactions are working simultaneously in the H- and He-burning shells which therefore advance at a similar rate. The intershell region thus does not grow in mass and thermal pulses are  inhibited.
Interestingly, \cite{chieffi1984} realised that an envelope pollution as low as $Z_\mathrm{surf}\sim 10^{-6}$ was enough to reactivate the occurrence of thermal pulses. 

These results were accompanied and supported by the work of \cite{fujimoto1984}. 
They developed a semi-analytical model to study the general behaviour of the thermally-pulsing AGB stars of the lowest metallicities. They considered the degenerate core mass and the envelope metallicity as key parameters of their analysis. It was established that stars hosting pristine envelopes drastically changed their behaviour when the core mass reached a critical value of M$_1^*$ = 0.73~\msun. 

This critical core mass corresponds to the transition between a H burning shell powered by the pp chains (in low mass stars) and the CNO cycles (in more massive stars). Stars with core masses below M$_1^*$ are able to undergo He-shell flashes, whereas those with more massive degenerate cores develop steady He-shell burning.
Actually, above M$_1^*$ the occurrence of thermal pulses depends on the envelope composition. As demonstrated  by \cite{fujimoto1984}, if the CNO envelope mass fraction exceeds $X_\mathrm{CNO} \sim 10^{-8}$ 
then He shell flashes are present again. In the absence of (self-) pollution, it is therefore expected that most primordial intermediate-mass stars will end their lives as supernovae. We will develop this point further in Section \ref{sec:fates}.

The existence of thermal pulses in primordial stars was revisited by \cite{fujimoto2000},  \cite{dominguez2000} and \cite{chieffi2001}. Unlike expectations from former works, these authors did obtain thermal pulses for stars of initial mass between 5 and 8~\msun. Shortly afterwards \cite{siess2002}  presented `normal' thermally-pulsing AGB stars of primordial metallicity. The reason for this behaviour is explained with further detail in the following subsections. Here we just mention that it is related to an increase in surface metallicities ($Z_\mathrm{surf}\gtrsim10^{-6}-10^{-5}$), either during the early AGB (E-AGB), or during the first He-burning shell instabilities, and thus the essential physics of the result by \cite{fujimoto1984} and \cite{chieffi2001} still remained.

Later works by \cite{suda2004}, \cite{lau2008,lau2009} \cite{campbell2008} and \cite{suda2010} on the evolution of primordial and very metal-poor stars confirmed the occurrence of thermal pulses. \cite{gilpons2007} showed that, even after an extremely inefficient second dredge-up, which led to surface CNO abundances $\sim 10^{-9}$, thermal pulses still occurred. Therefore primordial stars do experience thermal pulses, even when their envelopes are just barely polluted during their E-AGB phase.

\subsubsection{Evolution as `normal' thermally-pulsing AGB and Super-AGB stars}

We have seen that `normal' thermal pulses follow if the core mass is lower than a critical value, or if the stellar envelope has been enriched in metals above some critical amount. This enrichment can arise from a previous dual core flash episode, an efficient second dredge-up episode, or the occurrence of mixing events at the beginning of the AGB or Super-AGB phase. This then leads to more or less efficient third dredge-up and/or hot bottom burning, and the activation of relatively strong stellar winds, which eventually allow the ejection of stellar envelopes. Then we may say that such metal-poor stars behave as `normal' thermally-pulsing AGB and Super-AGB stars. Here we describe the conditions for the occurrence of a `normal' thermally-pulsing AGB or Super-AGB phase in primordial to EMP stars.

\begin{figure}[t]
\begin{center}
\includegraphics[width=1\linewidth]{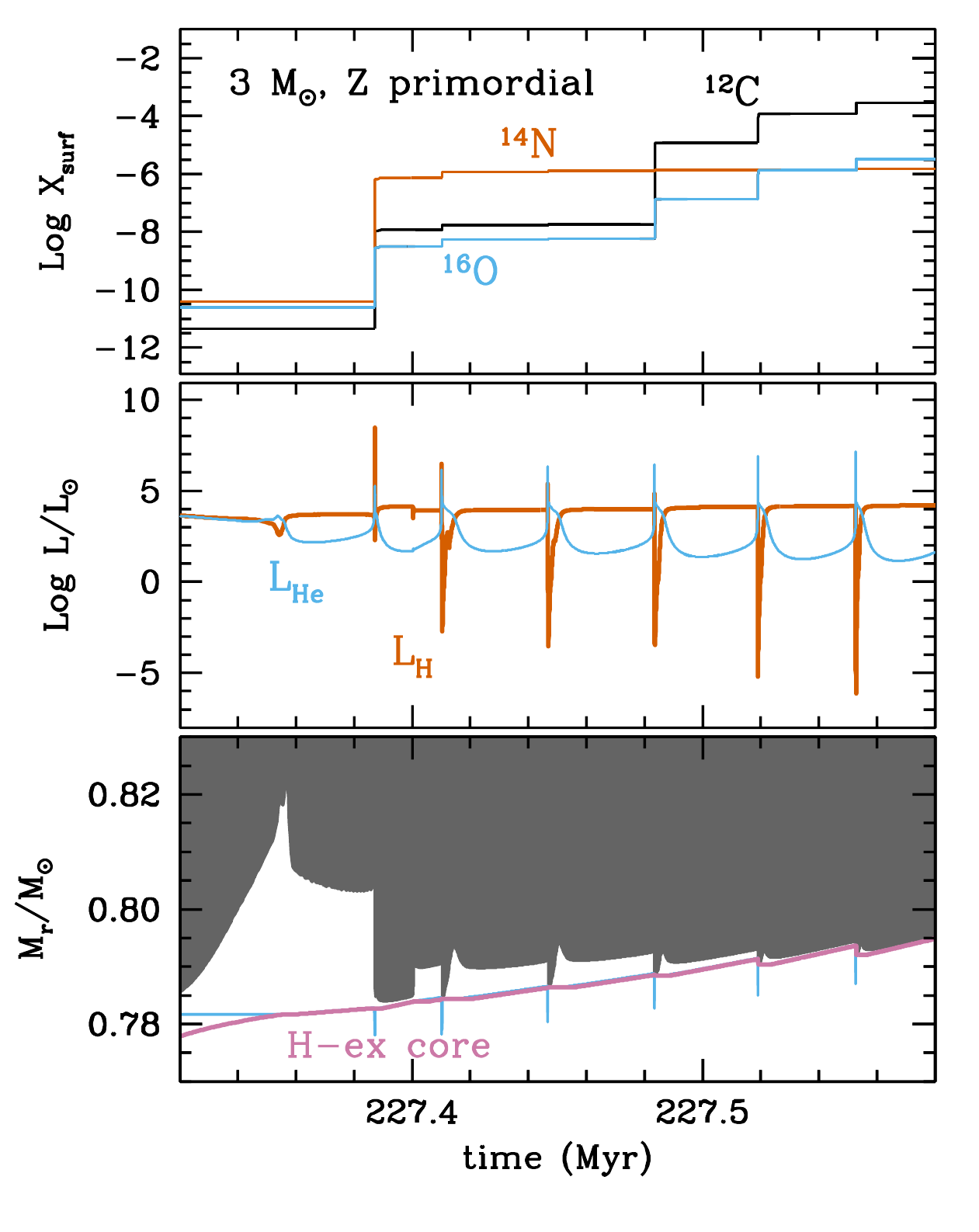}
\caption{First thermal pulses and dual shell flashes (DSF) of the thermally-pulsing AGB phase of the 3~\msun{} primordial model. Lower panel: evolution of the convective envelope (grey) and of inner convective shells (blue), as well as the evolution of the mass of the H-exhausted core (purple). Middle panel:  evolution of the luminosities associated with H and He burning (${\rm L_H}$ in blue and ${\rm L_{He}}$ in orange, respectively). Upper panel: evolution of surface abundances of C (black), N (orange) and O (blue).}
\label{fig:degpulses}
\end{center}
\end{figure}

\begin{itemize}

\item[-] {\it Dual Shell Flash and C-ingestion events:}

Models of initial mass $\rm 0.8~M_\odot \lesssim M_{ZAMS} \lesssim 1.3~M_\odot$ and metallicity below $\sim 10^{-6}-10^{-5}$ 
may experience one or several proton ingestion episodes (PIEs) during the thermally-pulsing AGB phase. 
These PIEs are similar to the DCF briefly outlined in Section \ref{sec:evolution}: in a DSF the low entropy barrier near the active burning regions allows the inner He-convective shell to extend upwards, beyond the limits of the H-exhausted core. 
This triggers a H-flash and the development of a small convective zone (see Figure \ref{fig:mixing}) enriched in carbon that later will be engulfed in the envelope, leading to its metal enrichment.
This phenomenon was studied in detail with 1D hydrostatic codes by, e.g. \cite{fuj90,fujimoto2000}, \cite{siess2002}, \cite{suda2004}, \cite{campbell2008}, \cite{iwamoto2009}  and \cite{suda2010}.  
However, as described in \cite{woodward2015} 
a correct investigation of these phenomena requires 
3D hydrodynamics with high spatial and temporal resolution. 
\cite{campbell2008} found that these DSF events occurred for initial masses $\rm 0.8~M_\odot \lesssim M_{ZAMS} \lesssim 1.3~M_\odot$.

Another PIE occurs at the beginning of the thermally-pulsing AGB phase for stars with masses $\gtrsim$ 1.3~\msun. In this case, following the development of an early pulse, a convective zone forms in the H-rich shell and extends inward to penetrate into the C rich layers. This process was analysed by \cite{chieffi2001}, who named it C ingestion.
As we saw with the DCF, the nomenclature for these phenomena is quite  heterogeneous. In Figure \ref{fig:mixing} we present the schematic behaviour of convective zones during DSF and C-ingestion episodes, and show the different nomenclature used to refer to these phenomena. 
Note that \cite{campbell2008} also use the term DSF to refer to PIEs that are initiated during a shell flash in stars of  $\rm M_{ZAMS} >$ 1.3~\msun. It should be noted that more metal-rich low-mass star models with $Z=10^{-4}$ have been reported to experience proton-ingestion episodes without the occurrence of dual flashes \citep{lugaro2012}.

The occurrence of DSF or C-ingestion episodes always leads to surface enrichments up to values $Z_\mathrm{surf} \sim 10^{-4}-10^{-3}$.
As a consequence, thermal pulses become stronger and stellar winds reach values more similar to those of higher metallicity thermally-pulsing AGB stars.
As an example Figure \ref{fig:degpulses} shows the evolution of a primordial 3~\msun{} star during the E-AGB and the first six thermal pulses. After a weak He-pulse the star develops four consecutive DSFs, that are able to highly enrich the stellar envelope in C, N and O. Later on, this model star continues its evolution similarly to a higher $Z$ object of the same mass:~it experiences the third dredge-up and ends its life as a white dwarf. 
It must be highlighted that DSFs may occur after the ignition of several mini-pulses or He-burning instabilities, which are too weak to allow for the formation of inner convective shells. This was the case reported by \cite{chieffi2001} and \cite{siess2002} for their 4 and 5~\msun{} primordial metallicity models respectively.

\vspace{3mm}
\item[-] {\it Efficient third dredge-up:}

\begin{figure*}[t]
\begin{center}
\includegraphics[width=0.80\linewidth]{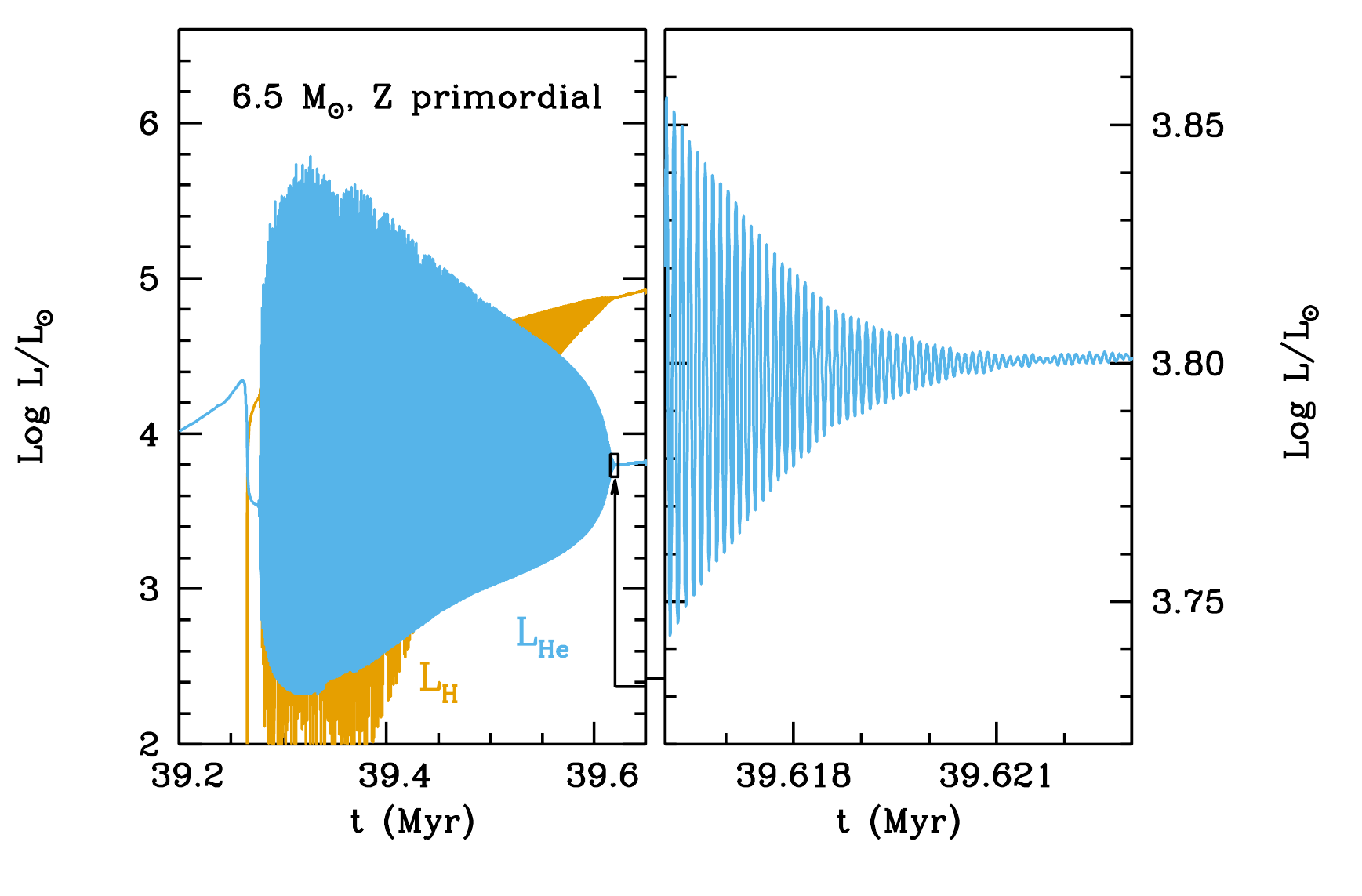}
\caption{Left panel: H- and He-burning luminosities ($\rm L_{H}$ in orange, and $\rm L_{He}$ in blue, respectively) during the thermally-pulsing AGB phase of a 6.5~\msun{} star of primordial composition. Right panel shows a zoom of the last thermal pulses represented on the left.}
\label{fig:tpagb65ms}
\end{center}
\end{figure*}

As reviewed  in the previous section, the occurrence of thermal pulses in EMP stars with core mass $M > \mathrm{M}_1^*$ depends on the metal content of the envelope. However, the ability of these pulses to drive a third dredge-up episode depends sensitively
on the treatment of convective boundaries.  
The primordial metallicity intermediate-mass models from \cite{chieffi2001} and \cite{siess2002} were calculated using overshooting. 
In particular \cite{siess2002} presented results with difussive overshooting, as proposed in \cite{freytag1996} and \cite{her97}. \cite{chieffi2001} and \cite{siess2002} reported efficient third dredge-up with positive feedback, which caused even further envelope pollution, stronger thermal pulses and thus even more efficient third dredge-up. As a consequence relatively strong stellar winds were expected from their models. 

The behaviour at somewhat higher metallicity ($Z \sim 10^{-6}$ and $Z \sim 10^{-5}$) is also strongly model dependent. \citet{gilpons2013} and \cite{lau2008} obtained efficient third dredge-up without including overshooting. Note however that the \citet{gilpons2013} models use the algorithm devised by \cite{frost1996} to determine the convective boundaries.
On the other hand \citet{suda2010}, using the strict Schwarzschild criterion, did not report any third dredge-up between 5 and 7~\msun{} approximately in the same metallicity regime.

\vspace{3mm}
\item[-] {\it Corrosive second dredge-up and Dredge-out:} 

\begin{figure}
\begin{center}
\vspace{-0.3cm}
\includegraphics[width=1\linewidth]{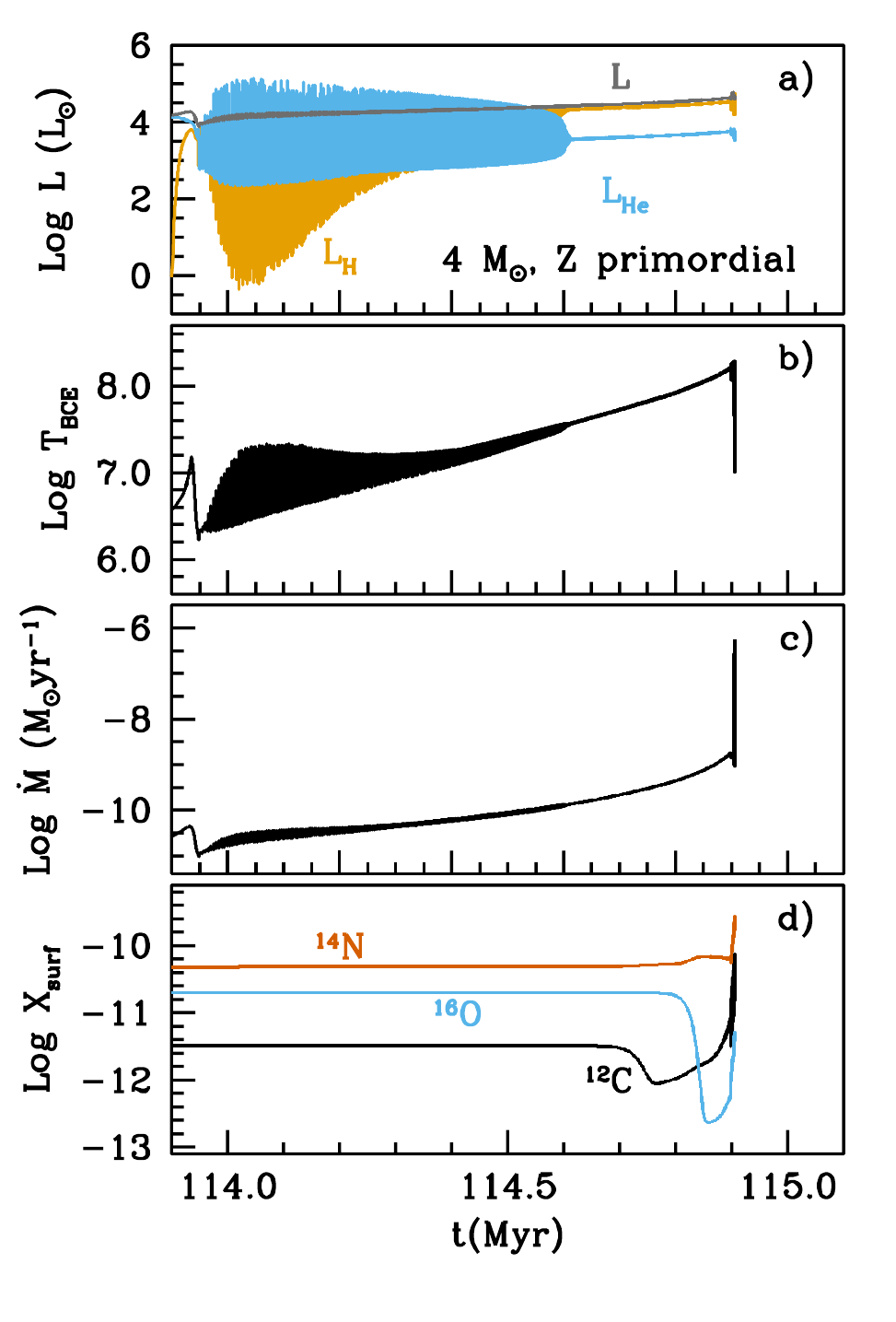}
\caption{ 
Summary of the evolution during the thermally-pulsing AGB phase of the 4~\msun{} primordial metallicity model. Panel a) shows the evolution of H- and He-burning luminosities ($\rm L_H$ in orange and $\rm L_{He}$ in blue), and the surface luminosity (L) in grey. Panel b) shows the evolution of the temperature at the base of the convective envelope. Panel c) shows the evolution of mass-loss rates, and Panel d) shows the evolution of surface abundances of $^{12}$C (black), $^{14}$N (orange) and $^{16}$O (blue).}
\label{fig:tpagb4ms}
\end{center}
\end{figure}

		Primordial to $Z=10^{-8}$ stars of initial mass 7~\msun{} $\lesssim  \rm M_{ZAMS} \lesssim$ 9~\msun{} experience a corrosive second dredge-up prior to the thermally-pulsing Super-AGB phase, and third dredge-up episodes later on. Therefore their stellar envelopes are enriched in metals (specially C and O) and, again, their evolution is more similar to that of `normal' thermally-pulsing Super-AGB stars. Mass-loss rates during the thermally-pulsing Super-AGB for stars with $\rm M_{ZAMS} \gtrsim$ 9~\msun{} are even higher ($\dot M \sim 10^{-5}$~\msun{} yr$^{-1}$) as a consequence of the dredge-out episode. 
\end{itemize}

\subsubsection{The cessation of thermal pulses} \label{sec:cessation}

The occurrence of the second dredge-up is not enough to ensure a standard thermally-pulsing AGB or Super-AGB behaviour in intermediate-mass stars. One of the most interesting and peculiar features of primordial thermally-pulsing AGB and Super-AGB stars was presented by \cite{lau2008}. These authors described the decrease in the intensity and the eventual disappearance of thermal pulses in primordial 5 and 7~\msun{} models. 
Their results can be explained by the narrowing of the  He-rich intershell, which reduces the amount of fuel 
and by the higher temperature of the intershell that increases the contribution of radiation to the total pressure and make in this regime the 3-$\alpha$ reaction rate  less dependent on temperature \citep[e.g.][]{siess2007}. As a consequence, the thermal pulses are weaker and the corresponding expansion much more moderate than for higher metallicity stars \citep[see][for a detailed analysis of the stability criteria]{yoon2004}.

The results for a similar calculation are presented in Fi\-gure \ref{fig:tpagb65ms}, for a primordial 6.5~\msun{} model, and in Figure \ref{fig:tpagb4ms}, for a 4~\msun{} model. In both cases we find, as did \cite{lau2008}, that our thermal pulses decrease in intensity and eventually disappear. Later on both H- and He-burning proceed quiescently,  
but other interesting evolutionary events 
are encountered (Guti\'errez et al., 2018, in preparation):~a few $10^4$ years after the disappearance of thermal pulses, when the core mass is $\sim 1.05$~\msun, and the temperature at the base of the convective envelope reaches $100\times 10^6$ K,
the 3$\alpha$ reactions are also activated at the base of the convective envelope, which causes a mild increase of $^{12}$C at the stellar surface, even when no third dredge-up is active. This increase in envelope metallicity may eventually boost unstable He-burning, and trigger third dredge-up if, as expected by \cite{kom07}, this phenomenon happens above a critical $Z$. Therefore at this point, the possibility of reaching a critical metallicity, as proposed in \cite{fujimoto1984}
cannot yet be discarded for models which experience the re-onset of thermal pulses. 
This might drive a new series of stronger thermal pulses and a significant envelope enrichment in carbon which, itself, might drastically enhance the mass-loss rates. 
It is interesting to note that the phenomena of the cessation and re-onset of thermal pulsations, with a different anatomy from standard thermally-pulsing AGB pulses, 
is also encountered with the code \textsc{mesa} (see \cite{paxton2018} and references therein). These new thermal pulsations have luminosities which, even at their local maximum values, are about one order of magnitude lower than the luminosity from H-burning, which also develops through pulsations (see Figure \ref{fig:tpagb4ms}).
According to our calculations the range of masses for which primordial stars are expected to develop thermal pulses 
and end (or temporarily halt) them is between $\sim 4$ and $\sim 7$~\msun, when using the stellar wind prescriptions by either \cite{vassiliadis1993}, or \cite{bloecker1995} with $\eta=0.01$. Stars of $Z=10^{-8}$  proceed through the thermally-pulsing AGB or Super-AGB phase in a way very similar to that of primordial objects, that is, they also experience the end of thermal pulses, but in a narrower mass range (between $\sim 5$ and $\sim7$~\msun).

\subsubsection{Evolution as a function of mass and metallicity} 
\label{sec:regions}

Figure \ref{fig:evoregions} summarizes the expected main characteristics of the late evolutionary stages of stars between 3 and 10~\msun, from approximately primordial $Z$ to $\log Z=-3.5$. These results correspond to a set of calculations obtained with similar versions of the same code (\textsc{monstar}), and using similar input physics. It must be noted that the inclusion of different input physics, especially very different mass-loss rates due to stellar winds, different definitions of the convective boundaries, or fast rotation, would alter the locations of 
the quoted  regions. 
For instance, the limits of the different evolutionary regions proposed by \cite{fujimoto2000}, \cite{suda2004} and \cite{suda2010} do not coincide with the ones shown in Figure \ref{fig:evoregions}, but the existence of these regions and their dependence on initial mass and metallicity are reproduced. In particular, \cite{suda2010} find a wider initial metallicity interval in 
which no third dredge-up is occurring, probably because they used the strict Schwarzschild criterion (with no modifications) for their calculations. Even though they did not follow the advanced thermally-pulsing AGB or Super-AGB phase, we could expect that such models would end up experiencing a cessation of thermal pulses (our grey region). 
On the other hand, according to the results from \cite{chieffi2001} and \cite{siess2002}, which implemented overshooting, the grey area corresponding to the cessation of thermal pulses would probably disappear.  The reason is that their models experience third dredge-up, stronger thermal pulses, and overall, a thermally-pulsing AGB or Super-AGB phase more similar to that of higher $Z$ stars.

\section{THE MAIN INPUT PHYSICS AND MODEL UNCERTAINTIES} \label{sec:uncertainties}

\subsection{The efficiency of third dredge-up} \label{sec:uncert-tdu}

The correct determination of convective boundaries is critical in many stages of stellar evolution. Here we focus on the third dredge-up, which is of prime importance for the evolution and fates of the lowest metallicity intermediate-mass stars.

The efficiency of the third dredge-up is a long standing unknown in thermally-pulsing Super-AGB calculations. Regardless of the initial metallicity, the third dredge-up is intimately related to the treatment of convective boundaries. Models which implement the strict Schwarzschild criterion either experience a less efficient or no third dredge-up at all (\citealt{siess2007}, \citealt{gilpons2007}, \citealt{lau2008}). On the other hand, models that either implement a modification of the Schwarzschild limit, such as the attempt to search for convective neutrality  (see \cite{frost1996} and the discussion at the beginning of section \ref{sec:evolution}), or overshooting (\citealt{herwig1999}, \citealt{chieffi2001}, \citealt{siess2002}) usually find efficient third dredge-up (see, for instance \citealt{herwig2000}, \citealt{her04a}, \citealt{cristallo2009}, and \citealt{karakas2010}). 

The efficiency of the third dredge-up also depends on the strength of the thermal pulses, because strong pulses drive further expansion and cooling of the regions below the base of the convective envelope. This cooling increases the opacity and thus produces a deeper inward progression of convection.

At least for relatively low mass and higher metallicity objects 
the effects of the third dredge-up on surface composition can be compared with observations, and thus allow some calibration (e.g. \citealt{marigo1999}, \citealt{girardi2003}). In the case of EMP stars, the occurrence of third dredge-up can be derived from the presence of $s$-process elements in the surface of unevolved C-enhanced EMP stars.
The difficulty in reliably determining the third dredge-up efficiency limits our knowledge of the final fates, since the third dredge-up not only alters the metal content of the envelope, but also determines the core growth rate\footnote{A large amount of overshooting at the boundaries of He-flash driven convective zones may lead
to a decrease in CO core size and to an enhancement in third dredge-up efficiency \citep{herwig2000}. Whether this effect is real remains to be determined.}, and the mass-loss rates due to stellar winds. \cite{her04a}, \cite{goriely2004}, and \cite{lau2009} reported the occurrence of a `hot third dredge-up', which occurs at envelope temperatures so high that some C may be transformed into N during the process. 
During a hot third dredge-up the convective envelope is able to erode most of or, in some cases, even the entire intershell, and reach the CO core.  Furthermore, the depth of third dredge-up determines the composition of the envelope which determines the local opacity, which feeds back onto the depth of dredge-up.  The envelope composition also has a substantial effect on the mass-loss. We will consider it in subsection \ref{sec:mloss}.

Finally, it is important to recall the relevance of numerics in these evolutionary calculations. As reported by \cite{chieffi2001}, changing the time step or spatial resolution may affect the advance of the convective envelope into C rich regions.

\begin{figure*}[t]
\begin{center}
\includegraphics[width=0.70\linewidth]{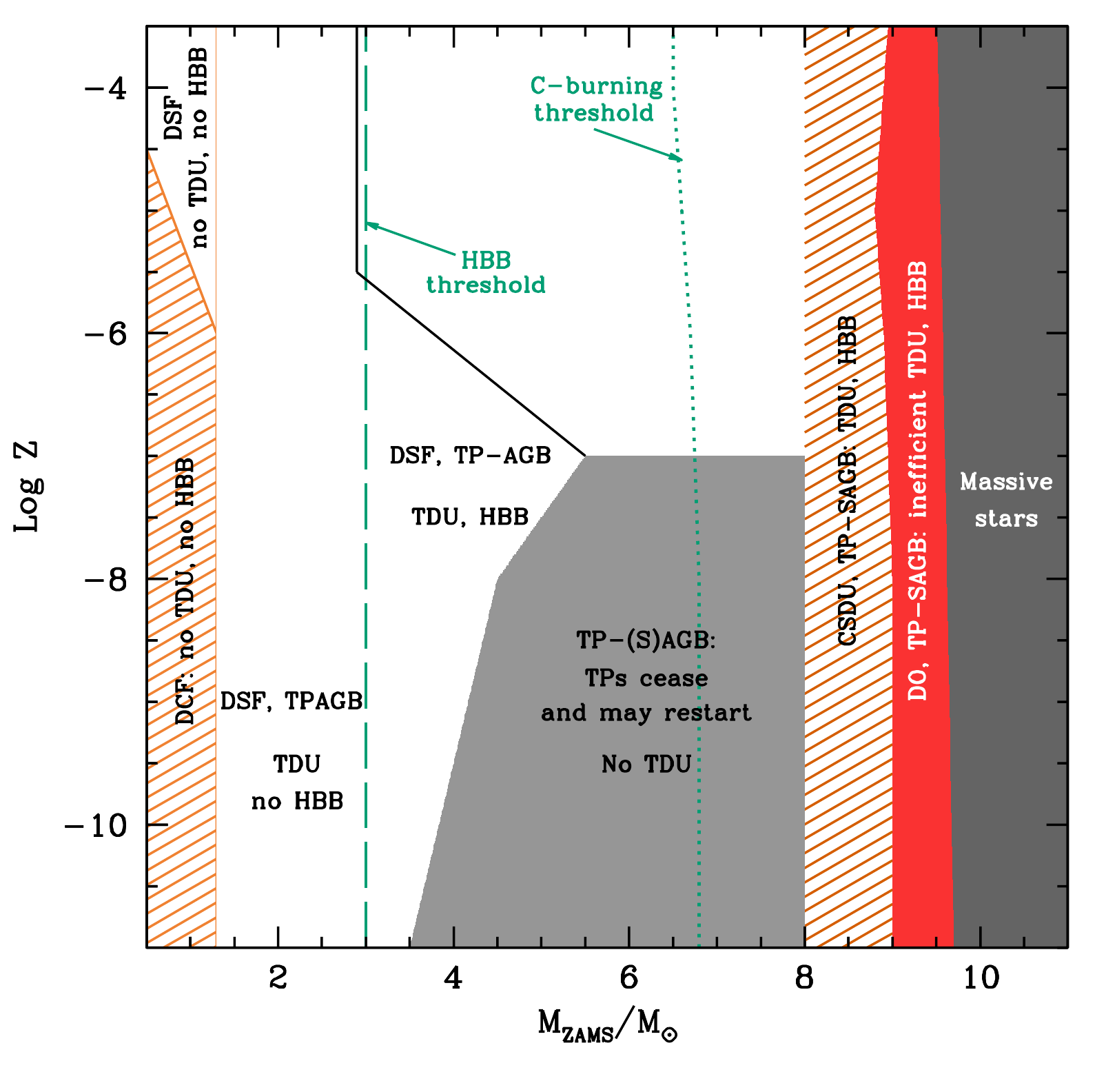}
\caption{Approximate classification of primordial to very metal-poor models in the $\rm M_{ZAMS}$--log $Z$ plane, according to the main characteristics of their late evolution.
Models to the right of the green dotted line experience C burning. Models to the right of the green dashed line experience hot-bottom burning (HBB). 
DCF refers to Dual Core Flash, DSF to Dual Shell Flash, DO to dredge-out, TDU to third dredge up, and CSDU to Corrosive Second Dredge-Up. See text for further details.}
\label{fig:evoregions}
\end{center}
\end{figure*}

\subsection{The effect of different sources of opacities: molecular opacities and dust}

In the low-temperature regime (T $\lesssim$ 5000 K) molecules and dust are the main sources of opacity.
Low-temperature opacities were traditionally calculated under the assumption of a scaled solar composition (see, for instance \citealt{alexander1975} and  \citealt{ferguson2005}), and thus could not account for the envelope abundance variations caused by second and third dredge-up episodes, and by hot bottom burning. This important drawback was alleviated either by interpolating within existing opacity tables to account for the CN molecule \citep{scalo1975}, or variable C abundances, \citep{bessell1989}, or by calculating new opacity tables with variable C/O ratios, such as \cite{alexander1983} and \cite{lucy1986}. 

The effects of variable composition low-temperature opacities in evolutionary calculations were highlighted by the synthetic models of \cite{marigo2002}, and then in the detailed AGB models of 
\cite{cristallo2007}, 
\cite{weiss2009}, 
\cite{ventura2009},
\cite{ventura2010}, 
\cite{fishlock2014} and 
\cite{constantino2014}. 
The latter authors used the opacity tables in \textsc{aesopus} (\citealt{lederer2009}, \citealt{marigo2009}) and concluded that, regardless of their original metallicity, all model calculations of initial mass $\lesssim$ 3~\msun{} should include changes in the surface composition and their effect on opacity because, even at very low metallicities, models were able to efficiently dredge-up metals to the surface and significantly alter their surface composition. 
In general the consequences of including variable composition low-temperature 
effects include higher opacity values, larger radii, lower surface temperatures and higher mass-loss rates. As a consequence the thermally-pulsing AGB or Super-AGB phase is shorter, the third dredge-up is less efficient (there are fewer thermal pulses) and hot bottom burning is less efficient (when it occurs). 

Until very recently dust in the most metal-poor AGB stars was assumed to be practically non-existent \citep{dicrescienzo2013}, and thus an almost irrelevant source of opacity compared to molecules. However recent work by \cite{tashibu2017} suggests that dust might form after envelope pollution caused by the second dredge-up, by proton-ingestion episodes and by the third dredge-up. This additional source of opacity would further increase the effects of the composition-dependent molecular opacities as stated above.

It must be noted that for stars with $Z \lesssim 10^{-8}$ and initial masses 
$\rm 5~M_{\odot} \lesssim M_{ZAMS} \lesssim 8~M_{\odot}$, that neither undergo a very efficient second dredge-up, nor proton-ingestion episodes, nor a third dredge-up, the photosphere is too hot to allow for the formation of carbon dust which, according to \cite{tashibu2017}, occurs for T$_{\rm eff} \lesssim 3850$ K.

\subsection{Mass-loss rates}\label{sec:mloss}

A very substantial source of uncertainty, which compromises our knowledge of the final fate of the most metal-poor stars, is represented by stellar winds. It is known that intermediate-mass stars of `normal' metallicity lose their envelopes during their RGB and (super-)AGB phases to become white dwarfs. The exceptions to this general behaviour are the most massive intermediate-mass objects, whose outcome may be  either a white dwarf or an electron-capture SN. 
The situation is much more uncertain in the case of EMP stars. In ge\-neral, stellar winds are controlled by different mechanisms, such as radiation, pulsations and dust formation, or photospheric Alfv\'en waves, but a clear, self-consistent theory is still lacking. During the RGB, the standard choice was \cite{reimers1975} for a long time, but its shortcomings (related to the mechanical energy flux in the envelope, and to its dependence on the chromospheric height) prompted a revision of this prescription, which was addressed by \cite{vanloon2005}, \cite{schroeder2005}, and \cite{mcdonald2015}. With the new prescription by \cite{schroeder2005}, stellar winds agree with observed RGB mass-loss observations over a wide range of metallicities  (see \citealt{schroeder2007}).

The driving mechanism of stellar winds during the E-AGB may still be well described by \cite{schroeder2005}, but when the superwind phase ($\dot M \gtrsim 10^{-5}$~\msun{} yr$^{-1}$) is reached during the thermally-pulsing AGB, then alternative prescriptions based on pulsation aided dust driven winds must be considered. \cite{vassiliadis1993} established a direct relation between mass-loss rate and pulsation period after compiling CO microwave observations of AGB stars.
\cite{straniero2006} proposed a new calibration for the mass-loss period relation,
which gave results more similar to the prescription of \cite{reimers1975}, with a multiplying
 constant which switched from 0.5 to 5 on the late thermally-pulsing AGB. 
\cite{bloecker1995} presented a prescription based on the atmospheric calculations for Mira stars made by \cite{bowen1988}. The mass-loss rates derived from these different approaches differ widely, with \cite{bloecker1995} rates being far higher than the rest (by a factor $\sim$ 100). We note that most calculations which use Bloecker's prescription 
(even in works by Bloecker himself) tend to apply a multiplying constant $\eta \sim$ 0.01 (see, for instance, \citealt{ven10a}), or $\eta \sim$ 0.1, as in \citet{groenewegen1994}. 

Mass-loss rates associated with pulsations in the case of the most metal-poor stars present two main problems.  First, according to the traditional perspective, pulsations in AGB and Super-AGB stars are induced by radiation pressure in dust grains which, in principle, are absent (or existing only in small amounts) in the lowest-$Z$ cases. 
Dust around stars can be produced in either carbon-rich, or oxygen-rich chromospheres.  
Carbon is obviously required to form carbonaceous dust. This element can be both primary and produced in AGB stars (although not efficiently in some EMP stars). O, Si, Al and Fe are required for dust production in O-rich environments, but substantial amounts of Si and Al cannot be produced in the most metal-poor AGB stars. Additionally dust formation requires relatively low temperatures, whereas the most metal-poor stars are more compact and hotter than their higher $Z$ counterparts.   
The second reason why mass loss is thought to be reduced at lower metallicity regimes is related to the pulsations themselves.  From the theoretical pulsation model predictions from \cite{wood2011a} it is expected that, in EMP AGB stars, the amplitude of stellar pulsations is lower, and hence strong pulsation driven winds are inhibited.

Interestingly, none of the wind rate prescriptions mentioned above has an explicit dependence on metallicity. Of course the metallicity indirectly affect the mass-loss rates through its effect on surface luminosity, radius and effective temperature. Influenced by considerations related to stellar winds of more massive (and hotter) stars, a metallicity scaling ${(Z_\mathrm{surf}/Z_{\odot})^\alpha}$ was introduced by \cite{kudritzi1989}, where $Z_\mathrm{surf}$ is the stellar surface abundance, and $\alpha$ is an exponent typically ranging between 0.5 and 0.7. This scaling could account for the lower mass-loss rates expected from the most metal-poor stars, but its original justification was based on line-driven winds, which probably are not relevant for (super-)AGB stars, and limits its use to intermediate-mass stellar models.

As a consequence of the former considerations, the earliest works on advanced evolution of the most metal-poor stars assumed that stellar winds would be practically negligible. 
This apparently solid hypothesis was first shaken when detailed models showed that various mixing episodes were able to efficiently pollute stellar envelopes over a relatively wide mass range (see sections \ref{sec:prevmixing} and \ref{sec:uncert-tdu}). Later, when the composition dependent low-temperature opacities were introduced, stellar wind rates were dramatically enhanced, and the late evolutionary stages of intermediate-mass stars in the low-mass range, $\rm M_{ZAMS}$ $\lesssim$ 3~\msun, were shortened (\cite{constantino2014} and references therein).
Additionally, the possibility of forming dust in these stars also opened the possibility of very strong dust-driven winds as noted by \cite{tashibu2017}. These winds might cause the loss of the envelope in stars of initial mass below 
approximately 5~\msun.

Finally, because we expect low-temperature opacity effects to be less important in stars with 
$Z \lesssim 10^{-8}$ and initial masses $\rm 5\,M_{\odot} \lesssim M_{ZAMS} \lesssim 8\,M_{\odot}$, stellar winds in these objects could still be very low, and thus the characteristic thermally-pulsing AGB and Super-AGB evolution described in section \ref{sec:cessation}, with a thousand or more thermal pulses and their eventual disappearance is still expected.

\subsection{Additional sources of uncertainties}

\subsubsection{The instability in the late thermally-pulsing AGB and Super-AGB phase}

\cite{lau2012} analysed the reasons why thermally-pulsing AGB and Super-AGB model calculations fail to converge while their stellar envelopes are still relatively massive ($\rm M_{env}\sim$ 0.1-3~\msun). A sharp peak in the opacity, due to the presence of Fe-group elements, and located near the base of the convective envelope causes an accumulation of energy. This eventually leads to a departure from hydrostatic equilibrium and to the halting of calculations. The consequences of this instability are unclear: either the H-rich envelope might be quickly ejected, or hydrostatic equilibrium might be recovered after a fast envelope expansion. The lower Fe-peak element abundance in EMP stars might delay or hamper the occurrence of the instability, but this effect has not yet been studied in detail. 
\subsubsection{Nuclear reaction rates}
The most important reaction affecting the evolution of Super-AGB stars is $\rm ^{12}C(^{12}C,\alpha)^{20}Ne$. \cite{straniero2016} recently analysed the effects of taking into account an increase in this reaction rate, attributed to a possible resonance in the 1.3--1.7 MeV range that is expected from extrapolation of experimental data \citep{spillane2007}. According to \cite{straniero2016}, the effects of this modified reaction rate would be a decrease of $\sim$2~\msun{} in the lower initial mass threshold for C ignition, and a similar variation in the lower mass threshold for the formation of an iron-core leading to a core-collapse supernova. As a consequence, and regardless of the initial metallicity, the SN rate would be altered. These authors also analysed the effects of varying the important but highly uncertain rate of the $\rm ^{12}C(\alpha,\gamma)^{16}O$ reaction, but did not find significant effects on the mass thresholds mentioned above.

New experimental determinations of the rate of $\rm ^{12}C(^{12}C,\alpha)^{20}Ne$ and $\rm ^{12}C(^{12}C,p)^{23}Na$ by \cite{Tumino2018}
have reported an increase in the rate of $\sim 10$ over the standard rates by \cite{caughlan1988} in the range $0.5 - 1.2 \times 10^9$ K. These new rates, published in the late stages of the writing of this review, may have profound effects on the evolution of Super-AGB and massive stars, and change the initial mass thresholds for the different fates of stars.

\subsubsection{Rotation}

The  effects of rotation on the  evolution of intermediate-mass metal-poor stars has not been extensively studied but there is no reason to assume that it is not significant. In fact
metal-poor models are more compact and, thus, probably experience higher rotation rates than their higher metallicity counterparts (see, for instance, \citealt{mey07}, \citealt{ekstrom2008}). 
Hydrodynamical instabilities associated with meridional circulation and shear instability are expected to enhance mixing efficiency between the H-exhausted core and the envelope (\citealt{heger2000}, \citealt{maeder2001}, \citealt{meynet2002}, \citealt{chieffi2013}), especially at low metallicities. Therefore, it has important consequences in terms of nucleosynthesis. 

In terms of stellar final fates, it is important to consider that rotation may affect mass-loss rates  due to stellar winds \citep{heger2000}.
\cite{farmer2015}, found a very limited effect of rotation on the lower initial mass threshold for C ignition (at least when overshooting was included), although their analysis was restricted to solar metallicity models.  
\citet{decressin2009} computed intermediate-mass models with rotation in the metallicity range covered by globular clusters. They concluded that rotation favoured CNO surface pollution during dredge-up episodes, and thus higher metallicity ejecta during the thermally-pulsing AGB. 

Rotation affects many critical processes, such as mass-loss rates and transport of matter within stars. 
These transport mechanisms certainly interact with those already known to exist even in non-rotating stars. These facts led \cite{chieffi2013} to point out that a general solution to many discrepancies between observations and theoretical models might be found in a consistent treatment of rotation, rather than in separately tuning the effects of overshooting, or different mass-loss rate prescriptions.

\subsubsection{Binarity}
Many observed EMP stars belong to, or may be descendants of, stars that experienced binary interactions. Therefore it is important to highlight that a complete understanding of the evolution and nucleosynthesis of EMP stars should take these interactions into account. However, binarity can completely change the characteristics of the evolution and the fates of stars. Besides, the associated uncertainties add to (and are often entangled with) those of single EMP stars. A complete summary of the effects and uncertainties related to binarity would be a matter for a separate review and will not be discussed here.

\section{FINAL FATES OF PRIMORDIAL AND EMP STARS}\label{sec:fates}

The fate of stars that enter the thermally-pulsing AGB or Super-AGB phase depends on the competing effects of core growth and mass-loss rate by stellar winds. If the core is able to reach $\rm M_{Ch}$ before the envelope is lost, the star will become either a SN~I1/2 (\citealt{arnett1969}, \citealt{ibe83}) if it hosts a CO core, or an electron-capture SN (EC-SN), if it has an ONe core \citep{miy80, nomoto1984, nomoto1987}. If $\rm M_{Ch}$ is never reached, the star ends its life as a white dwarf. 
Both the core growth and mass-loss rates are based on the poorly known input physics described in Section \ref{sec:uncertainties}, which makes the determination of stellar final fates uncertain, especially at the lowest-$Z$ regime.

\begin{figure*}[t]
\begin{center}
\includegraphics[width=0.55\linewidth]{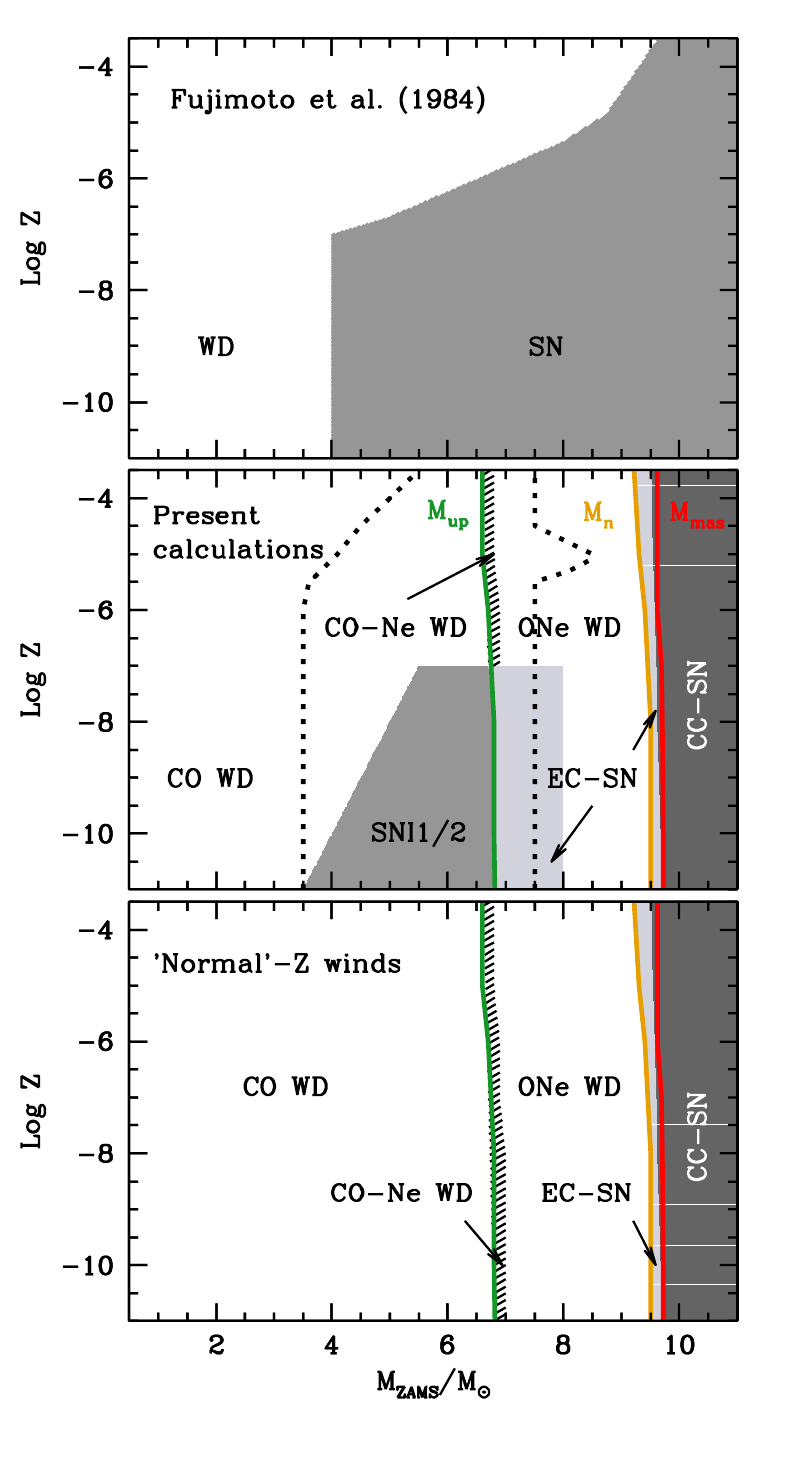}
\caption{Approximate regions defining the expected final fates for models of metallicity values between primordial and log $Z=-3.5$, in the initial mass--metallicity plane. Upper panels shows the expected final fates according to \cite{fujimoto1984}. The middle panel presents the final fates according to the evolution described in Figure \ref{fig:evoregions}. The region between the dotted lines represents the possible SN~I1/2 region derived from the work of \cite{suda2010}. The lower panel presents the
predicted final fates under the assumption that actual stellar winds in our models behave as those of `normal'  metal-rich stars. 
}
\label{fig:fates}
\end{center}
\end{figure*}

\subsection{The mass limits M$_{\bf up}$, M$_{\bf n}$ and M$_{\bf mas}$ as functions of the metallicity}

In discussing the final fates of intermediate-mass stars it is convenient to use the standard nomenclature:

\begin{itemize}
    \item $\rm{M_{up}}$: the minimum initial mass required to burn carbon sufficiently to develop an associated inner convective shell;
    \item ${\rm M_n}$: the minimum initial mass that leads to an EC-SN;
    \item $\rm M_{mas}$: the minimum initial mass that forms an core-collapse SN (see Figure \ref{fig:fates}). 
\end{itemize}

$M_{\rm up}$ is mainly controlled by the maximum size of the convective core during central H-burning and by the efficiency of the second dredge-up.
Different calculations, with different input physics and initial metallicities ranging between EMP and solar values, yield $M_{\rm up}$ values ranging between 5~\msun{}
(\citealt{tornambe1986}, \citealt{cassisi1993}, \citealt{gir00}), and 9~\msun{} \citep{siess2007}. The general trend with metallicity is the increase of ${\rm M_{up}}$ with $Z$, with a minimum ${\rm M_{up}}$ between $Z=10^{-4}$  \citep{siess2007} and  $Z=10^{-3}$ (\citealt{bec79}, \citealt{castellani1985},
\citealt{ume99}, \citealt{gir00}, \citealt{bono2000}, \citealt{ibeling2013}, \citealt{doherty2015}).

As shown in \cite{doherty2010}, models that are just above $\rm M_{up}$ ignite carbon in the very external shells of the CO core but the combustion quenches and cannot proceed to the centre. The stellar core then presents an atypical structure with a degenerate CO core surrounded by a thin layer of Ne and O. These failed Super-AGB stars develop so called hybrid CO-Ne cores and, according to \cite{doherty2015}, lie in a mass interval $\sim 0.1$~\msun{} wide above $\rm M_{up}$. This mass interval can increase to 1.4~\msun{} \citep{chen2014}, or even disappear \citep{brooks2016}, when different treatments of convective boundaries are implemented.

$\rm M_{mas}$ ranges between 8~\msun{} and 11.5~\msun{} \citep{poelarends2008} and its behavior as a function of metallicity is similar to that of $\rm M_{up}$. 
The mass interval between $\rm M_n$ and $\rm M_{mas}$ corresponds to the initial mass values over which EC-SNe form, and according to the latest calculations it is about 0.1--0.2~\msun{} wide \citep{doherty2015}. These results are in contrast to those from \cite{poe07}, who obtained an increasingly wide initial mass interval with decreasing $Z$ for the occurrence of EC-SNe, and the conclusion that all Super-AGB stars having $Z=10^{-5}$ would end their lives as EC-SNe.
The reason for these variations is the use of different input physics,
especially different prescriptions for the mass-loss rates. \citet{doherty2015} used the prescription by \cite{vassiliadis1993} with no additional dependence on the envelope metallicity. In contrast \cite{poe07} used the mass-loss prescription by \citet{vanloon2005} with the previously discussed metallicity scaling included.
In summary there are large variations in the different determinations of $\rm M_{up}$, $\rm M_n$, and $\rm M_{mas}$. This means that there are substantial
uncertainties in the initial mass interval for the occurrence of EC-SNe. This reflects the sensitivity of these quantities to uncertainties in the input physics and prescriptions for convection, and are at present unavoidable. 
Finally it should be noted that, whilst the final fates of stars with $Z \gtrsim 10^{-4}$ has been widely explored, only a few models at the lowest $Z$ regimes have been analysed.

\subsection{The formation of SNe~I1/2}

\citet{zij04} considered the reasoned assumption that stellar winds in the most metal-poor regime were very weak \citep{wood2011a}, and proposed that intermediate-mass stars with ${\rm M_{ZAMS} < M_{up}}$, i.e. those hosting CO cores during their thermally-pulsing phase, could become SNe~I1/2 \citep{arnett1969,ibe83}.

\cite{poe07} performed detailed calculations of intermediate-mass (and a few massive) stars up to the early AGB and Super-AGB, in order to obtain information about their envelope enrichment just after the second dredge-up and, 
especially, to get starting masses for their parameterised thermally-pulsing phase. This parametric approach was then used to analyse the subsequent model evolution and determine their final fates. 
The third dredge-up was parametrised as in \cite{kar02} and different prescriptions for mass-loss rates due to stellar winds were used 
(\citealt{vassiliadis1993}, \citealt{bloecker1995}, and \citealt{vanloon2005}). 
Their favoured parameterisation included the mass-loss prescription by \cite{vanloon2005} with an additional metallicity scaling from \cite{kudritzi1989}. 
Besides the occurrence of EC-SNe for all Super-AGB stars of $Z \sim 10^{-5}$ mentioned above, \cite{poe07} concluded that SN~I1/2 could form for initial masses between 6~\msun and 6.4~\msun, and that stars with $\rm M_{ZAMS} <$ 6~\msun{} would end up as CO-white dwarfs. These authors did not actually present detailed calculations below $Z\approx 10^{-5}$.

\cite{lau2008} presented calculations of the evolution of primordial 5 and 7~\msun{} models, whose thermal pulses lost strength and halted.
The 7~\msun{} model had experienced about 1400 pulses (see Section \ref{sec:tpagb}), and at the time of their cessation, it hosted a very low-metallicity envelope ($Z_\mathrm{surf}\sim 10^{-6}$). 
During the subsequent evolution, thermal pulses never recovered, and the degenerate core grew up to 1.36~\msun. At that point the star was still surrounded by a H-rich envelope and the physical conditions at the centre were very similar to those of a white dwarf belonging to a binary system just prior to a SN~Ia explosion. By analogy with SNe Ia, C burning under these conditions is not expected to lead to the formation of an ONe core but instead to the complete destruction of the star. This led the authors to conclude that their model of 7~\msun{} primordial star will produce a SN~I1/2.
 
The cessation of thermal pulses is found by various codes for models with $\rm M_{ZAMS}$ approximately between 4 and 7~\msun{} at primordial $Z$, and for models with $\rm M_{ZAMS}$ approximately between 5 and 7~\msun{} at $Z=10^{-8}$.

Using a parametric model to complement their detailed evolutionary calculations, \cite{lau2008} explored the possible outcomes of their models assuming a constant core-growth rate and different mass-loss rate prescriptions: specifically, \citet{reimers1975}, \citet{bloecker1995} and \citet{schroeder2005} both with and without metallicity scaling. The final fates of the considered stars were independent of the tested wind prescriptions, but were affected by the $Z$-scaling: a small $Z$-scaling expressed as $(Z/Z_{\odot})^{0.5}$ allowed the model to become a SN~I1/2.

The models presented by \cite{suda2010} also showed the existence of a region in the initial mass--initial metallicity plane where  third dredge-up does not develop (see Section \ref{sec:regions}). This fact together with the absence of a previous efficient second dredge-up allows us to infer that the expected final fate of these models might also be a SN~I1/2. The summary for the expected final fates according to different calculations (and input physics assumptions) is shown in Figure \ref{fig:fates}. It emphasizes the huge limitations in our knowledge of the fates of many EMP stars.

It is also important to realise that the calculations of models leading to the cessation of thermal pulses and, eventually, to the formation of SNe~I1/2, were performed without including composition dependent low-temperature opacities.
In principle it should not drastically alter these results, as the envelope metallicity at the onset of thermal pulses is very low ($Z_\mathrm{surf} \sim 10^{-6}$ in \textsc{stars}, $Z_\mathrm{surf} \lesssim 10^{-8}$ in \textsc{monstar}, and $Z_\mathrm{surf} \lesssim 10^{-7}$ in \textsc{mesa}). 
Besides, the recently found phenomenon of the re-onset of thermal pulses (Guti\'errez et al. 2018, in preparation) might completely change the picture concerning the occurrence of SNe I1/2. The reason is that, together with the new pulses, significant envelope enrichment and much more efficient winds could develop. This might prevent the core mass from reaching ${\rm M_{Ch}}$ before the envelope is completely lost.

The comparison between the former models (\citealt{lau2008}, \citealt{lau2009}) and the works by \cite{chieffi2001} and \cite{siess2002} illustrates the importance of the efficiency of the dredge-up episodes and, ultimately, of the treatment of convective boundaries.
thermally-pulsing Super-AGB models of intermediate-mass stars  presented by \cite{chieffi2001} and \cite{siess2002} that implemented diffusive overshooting show somewhat higher envelope metallicity after the second dredge-up and, most importantly, do experience an efficient third dredge-up. Thus they are able to drive stronger thermal pulses and moderately high stellar winds. Even though these authors did not follow the evolution until the end of the thermally-pulsing AGB or Super-AGB, one could reasonably expect that their model stars would end their lives as white dwarfs. 

In terms of applications of these models, \cite{matteucci1985} considered the effects of taking into account SNeI1/2 in galactic chemical evolution.
\cite{tsujimoto2006} interpreted the composition of low $\alpha-$ and n-capture 
element EMP stars in terms of the existence of SNI1/2 progenitors (they named these objects SNeIIIa).
\cite{suda2013} investigated the occurrence of SNeI1/2 in their analysis of the transition of the initial mass function using binary population synthesis.

\subsection{The formation of EC-SNe}\label{sec:ecsne}

Super-AGB stars whose ONe cores grow up to  $\rm M_{\rm core} = 1.37$~\msun{} (\citealt{miy80}, \citealt{nomoto1984}, \citealt{nomoto1987}) reach central densities high enough to make electron capture  reactions energetically favorable. In the ONe core, the electrons are captured by  $^{24}$Mg, $^{23}$Na, $^{20}$Ne and with a reduction of the electron density, the degenerate core loses its pressure support and starts to contract rapidly. Oxygen eventually ignites and the core is converted into a mixture resulting from nuclear statistical equilibrium. The subsequent electron captures  on these elements accelerates the collapse and a SN explosion supported by neutrino heating ensues \citep{kitaura2006}. The most massive Super-AGB models are also able to ignite Ne off-centre at the end of the C burning process. 
If the Ne-burning flame is quenched before reaching the centre, the star 
will also probably end its life as an EC-SN. The characteristics of Ne-burning in these peculiar stars strongly depend on the treatment of convective boundaries. The use of some convective boundary mixing may allow the occurrence of Ne-burning through a series of flashes which eventually get stalled and allow the formation of an EC-SN. Models undergoing this type of evolution have been named `failed massive' stars (\citealt{jones2013}, \citealt{jones2014}). On the other hand, when using the strict Schwarzschild criterion, the Ne-burning flame reaches the centre and the star continues its evolution to become an core-collapse SN (CC SN).

The  lower and upper initial mass thresholds for the formation of EC-SNe ($\rm M_n$ and $\rm M_{mas}$ respectively) for metallicities $\geq 10^{-5}$ were discussed in detail by \cite{doherty2017}. Here we focus on the most metal-poor cases
($Z \lesssim 10^{-5}$). It is interesting to note from the middle panel of Figure \ref{fig:fates} that there is a gap in the EC-SN region between 8~\msun{} and 
$\rm M_n$. That is, white dwarfs are expected to form in this mass range, even at the lowest metallicities.\footnote{We have artificially kept the notation $\rm M_n$ to refer to the minimum mass for stars which become EC-SNe `after undergoing a corrosive second dredge-up'. Strictly speaking, $\rm M_n$ also lies just above the upper limit for the formation of SNe~I1/2 in the primordial and $Z=10^{-8}$ cases. Our motivation for this choice of notation is the existence of a gap in initial mass for the formation of EC-SNe, and the continuity with the higher $Z$ cases.}
This gap in the EC-SN region is caused by the occurrence of the corrosive second dredge-up (see Figure \ref{fig:evoregions}), which pollutes the stellar envelope enough to allow for a `normal' thermally-pulsing Super-AGB, and thus for the occurrence of third dredge-up, moderately strong winds, and final fates as ONe WDs. 
The efficiency of third dredge-up, even though highly uncertain, is expected to decrease and become very low in the most massive intermediate-mass stars (in particular when $\rm M_{ZAMS} \gtrsim M_n$). As a consequence stars of initial mass above M$_{\rm n}$ may experience somewhat higher core growth rates on an initially massive core (close to $\rm M_{Ch}$) and then explode as EC-SNe. Between 6 and 8~\msun{} the absence of thermal pulses combined with a weak mass-loss rate allows the ONe core to reach the critical value of 1.37~\msun{} for an EC-SN.

In any case, the uncertainties in mass-loss rates at these metallicities are such that some exploration of different rates is required. A simple but useful way of doing this is the approach by \cite{siess2007}. This author defined the $\zeta$ parameter, the ratio of the average envelope mass-loss rates $ (\dot{M}_\mathrm{env})$ to average effective core growth rates $(\dot{M}_\mathrm{core})$ during the thermally-pulsing Super-AGB phase. i.e. $\zeta=\left|\frac{\langle{\dot{M}_\mathrm{env}}\rangle}{\langle{\dot{M}_\mathrm{core}}\rangle}\right|$. He demonstrated that the values of the critical masses $\rm M_n$ and $\rm M_{mas}$  depend only on this parameter and the core mass at the beginning of the thermally-pulsing Super-AGB phase. 
According to the detailed calculations by \cite{gilpons2013} for $Z=10^{-5}$, $\zeta \approx 73, 75$ and 220 for $\rm M_{ZAMS}= 7,\ 8$ and 9~\msun{} respectively. The latter value is considerably larger due to the high efficiency of the dredge-out in increasing envelope metallicity and ultimately driving high mass-loss rates. As a reference, considering a typical average core growth rate about $10^{-7}$~\msun{} yr$^{-1}$, values of  $\zeta \approx$ 75 and $\zeta \approx$ 220 would correspond to an average mass-loss rate of $7.5\times 10^{-6}$~\msun yr$^{-1}$ and $2.2\times 10^{-5}$~\msun yr$^{-1}$ respectively.  

\begin{figure}
\includegraphics[width=8.5cm]{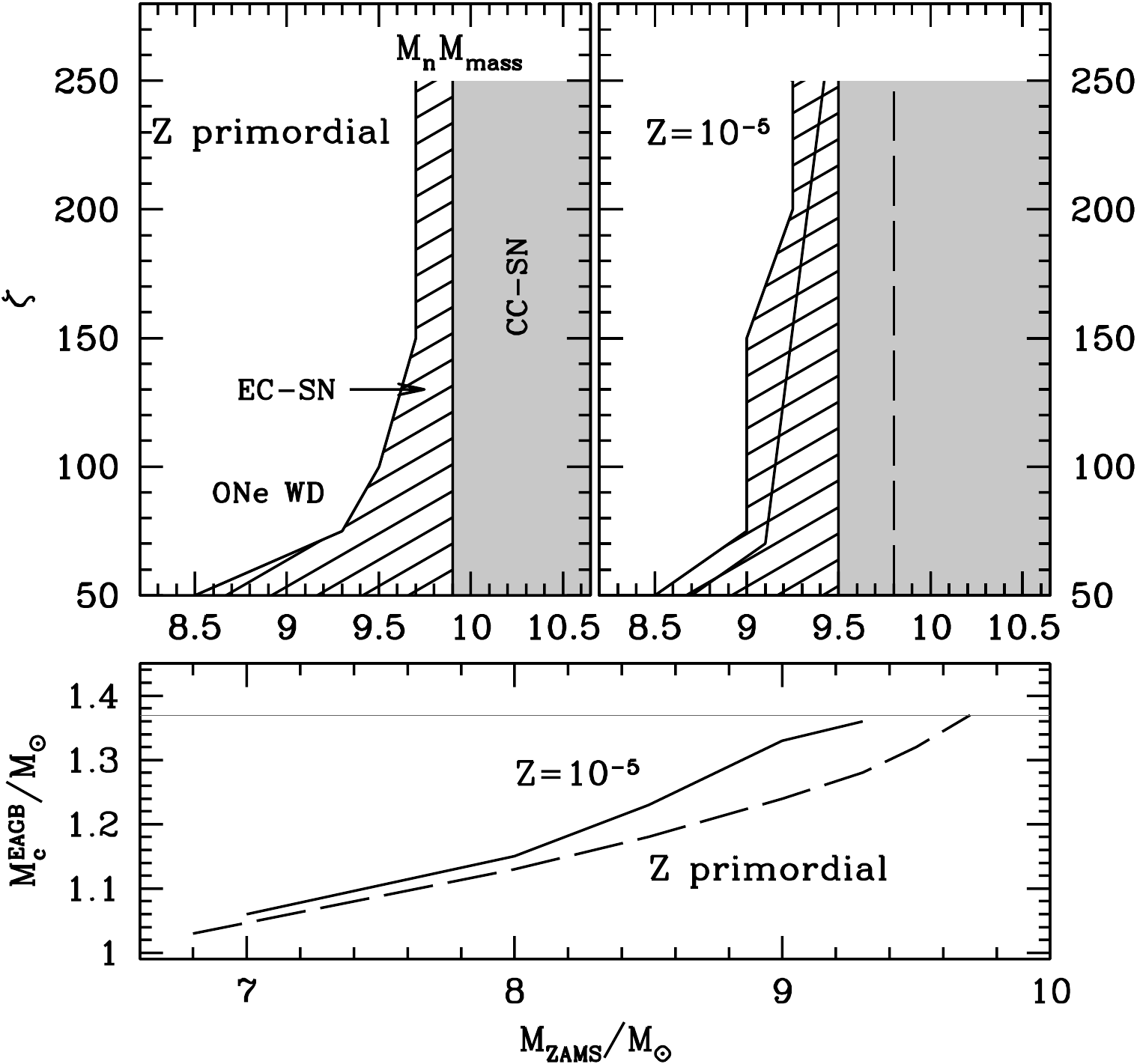}
\caption{
Lower panel: masses of the ONe degenerate cores versus ZAMS masses at the beginning of the thermally-pulsing Super-AGB phase for the primordial and $Z=10^{-5}$ cases. \cite{siess2007} results for $\rm M_n$ and $\rm M_{mas}$ at $Z=10^{-5}$ are shown in black solid and dashed lines respectively. Upper panels: expected fate versus initial mass for different values of the parameter $\zeta=\left|\frac{\langle{\dot{M}_\mathrm{env}}\rangle}{\langle{\dot{M}_\mathrm{core}}\rangle}\right|$ for the primordial cases (left) and the $Z=10^{-5}$ cases (right). }
\label{fatepoor}
\end{figure}

The evolution of $\rm M_n$ and $\rm M_{mas}$ as a function of $\zeta$ for the primordial and $Z=10^{-5}$ cases is illustrated in Figure \ref{fatepoor}. The interval of initial ZAMS mass that leads to the formation of EC-SNe in the primordial case ranges between 1.4~\msun{} for $\zeta=50$ (very slow winds) to 0.2~\msun{} for $\zeta \gtrsim$ 150. For the $Z=10^{-5}$ models we get wider ZAMS mass ranges, between 2 M$_{\odot}$ for $\zeta$ = 50, and 0.25~\msun{} for $\zeta \gtrsim$ 200. These intervals are similar (although shifted to somewhat lower initial masses) to the ones obtained by \cite{siess2007}. 
It is important to recall that uncertainties related to the treatment of convective boundaries and mass-loss rates affect the width of the ${\rm M_{ZAMS}}$ interval for the formation of EC-SNe, regardless of their initial metallicity. We refer the interested reader to \cite{jones2013} and \cite{doherty2017} for analyses of these effects.

\section{OBSERVATIONS OF EMP STARS}\label{sec:observations}

Uncertainties in nucleosynthetic yields of the most metal-poor stars derive from the unknowns in their evolution which we described in Section \ref{sec:uncertainties}, and from the difficulties in obtaining observational constraints, at least by comparison with higher metallicity stars.
The sample of observed objects at the most metal-poor regime has increased significantly in the last decade. Currently about 500 stars have been detected with [Fe/H]$\leq-3$. However , the interpretation of these observations is hampered by the need of considering a number of unconfirmed hypotheses in terms of the nature and IMF of ancient stars, of the chemodynamical evolution of the early universe and, as discussed here, in terms of stellar evolution and nucleosynthesis.

Observational information relevant for the understanding of the most metal-poor stars can be gathered from different sources. 
{\it Galactic archaeology} (\citealt{freeman2002}, \citealt{cohen2002}, and \citealt{carretta2002}) aims to understand the formation and evolution of the Milky Way through systematic study of its stellar populations. {\it Dwarf galaxy archaeology} aims for the same goal by considering stellar populations within dwarf galaxies \citep{frebel2012}. In both
cases the associated stellar database is a treasure trove for understanding the stellar populations themselves, in addition to using them as tools for understanding galaxies. Finally far-field cosmology of damped Ly$\alpha$ systems provide us with additional information from the high redshift universe \citep{cooke14}.

Stars with $\rm [Fe/H] \lesssim -3$ (EMP stars) are indeed uncommon, and become very rare at the lowest metallicities. Despite the continuous observational efforts made in the last decades, only $\sim10$ stars are known to have [Fe/H] $\lesssim -4.5$, including the latest discoveries of stars with [Fe/H] $<$ -5 (see \citealt{bonifacio2018}. and \citealt{aguado2018}). These efforts continue (see Section \ref{sec:intro}), and will probably provide us with further data down to [Ca/H] about $-9.4$  \citep{frebel2015}.
This value represents the detectability threshold of the CaIIK line, which is the proxy for Fe when it cannot be detected because of its low abundance.  
The exclusive group of EMP stars display a number of interesting peculiarities. 
We refer to the recent review by \cite{frebel2015} for a detailed description of observational data for EMP stars, and here we provide a summary of some of the most salient features.
Among these features we find that:
\begin{itemize}
\item[a)] EMP stars display a statistically significant abundance scatter \citep{matsuno2017}. This scatter is larger at the lowest observed [Fe/H]. 
\item[b)] EMP stars display different kinematic and chemical properties depending on whether they belong to the inner or to the outer Galactic Halo (\citealt{carollo2007}, \citealt{carollo2012}, \citealt{lee2017}). The outer Halo has a lower [Fe/H] population than the inner one. The most metal-poor stars of the Galactic bulge also present peculiar characteristics, in particular lower C enrichments than halo components \citep{howes2015}.
\item[c)] The Spite Plateau \citep{spite1982}, that is, the practically constant Li abundance value (A(Li)=$2.05 \pm 0.16$) measured in warm metal-poor stars was initially assumed to be representative of the Li produced during Big-Bang nucleosynthesis. This hypothesis had to be discarded mainly for two reasons. First, Big-Bang nucleosynthesis calculations yield Li abundances about $0.4$\,dex above the Spite Plateau. Second, the Plateau fails at metallicities [Fe/H] $\lesssim -2.8$. Below this value Li abundances show a wide scatter in which the characteristic value of the Spite Plateau becomes just an upper threshold (\citealt{ryan1996}, \citealt{ryan1999}, \citealt{boesgaard2005}, \citealt{asplund2006}, \citealt{bonifacio2007}, \citealt{aoki2009}). 
\item[d)] There is a high occurrence of C-enriched objects, increasingly higher at the lowest metallicities\footnote{C enrichment corresponds to [C/Fe] $>$ 1 according to \cite{beers2005}, and to [C/Fe] $>$ 0.7 according to \cite{aok07}.}. About 30\% of stars below [Fe/H] $\sim-3$ are C enriched, and this proportion goes up to about 80\% for $\rm [Fe/H] \lesssim -4$ (\citealt{cohen2005}, \citealt{frebel2005}, \citealt{lucatello2006}, \citealt{yong13}, \citealt{placco2014}). 
Their abundance pattern motivated the use of the specific terminology C-enhanced EMP or CEMP stars to refer to them \citep{beers2005}. CEMP stars are further subdivided into CEMP-$s$ (with [Ba/Fe] $>$ 0), CEMP-$r$ (with [Eu/Fe] $>$ 0), CEMP-$r/s$ or CEMP-$i$, as discussed below (with [Ba/Fe]$>$0 and [Eu/Fe] $>$ 0), and CEMP-no (neither s- nor r-enriched).
\item[e)] CEMP-$s$ stars are very frequent at metallicities $\rm -3 \lesssim [Fe/H] \lesssim -2$, but become rarer below these values \citep{aok07}\footnote{Note the heterogeneous classification criteria for these objects. Different authors define CEMP-$s$ as CEMP stars with [Ba/F]$>$1 and/or [Ba/Eu]$>$0.5 (\citealt{jonsell2006}, \citealt{lugaro2009}, \citealt{masseron2010}, \citealt{lee2013}).} 
Currently the lowest metallicity for CEMP-s stars, discovered by \cite{matsuno2017}, is around [Fe/H] = -3.6.
\item[f)] CEMP-no stars seem to show higher O-enhancements than CEMP-$s$ stars, and the N content shows a bimodal distribution with two distinct groups characterized by a high and low N-enrichment \citep{frebel2015}. There might be a correlation between $^{12}$C/$^{13}$C and [C/N] in CEMP-no stars \citep{nor13}.
\item[g)] In contrast to C-normal stars, CEMP-no stars display large spreads (~2 dex) in light elements (Na,
Mg, Al). They also show a moderate spread in Si, while the spread is small in heavier elements such
as Ti and Ca (see \cite{aoki2018} references therein).
\item[h)] NEMP stars are N-enhanced EMP stars \citep{izzard2009,pols2012}, such that [N/Fe] $>$ 1 and [N/C] $>$ 0.5. They appear to be more frequent at [Fe/H]$\lesssim$-2.8.
\item[i)] EMP stars tend to be $\alpha$-enhanced, that is with enrichment in $^{16}$O, $^{20}$Ne, $^{24}$Mg, $^{28}$Si, etc, up to $^{40}$Ca and $^{48}$Ti. Note that $^{48}$Ti is technically an Fe-peak element, although it behaves like an $\alpha-$element in metal-poor stars \citep{yong13}.
\item[j)] Finally, it should also be noted that there are a number of EMP stars which do not seem to fit in any of the groups mentioned above \citep{cohen2013}.  
\end{itemize}

\section{NUCLEOSYNTHESIS IN EMP STARS}\label{sec:nucleosynthesis}

Observations of EMP stars help us constrain our knowledge of the primitive universe and, in particular, the IMF of the first stars, the characteristics of their evolution, their final fates and their nucleosynthetic yields. In this section we review our current knowledge of EMP nucleosynthesis and relate this information to the observational features described in section \ref{sec:observations}. Ultimately, our goal is to understand which observed features of EMP stars may be explained with different stellar models, considering their nucleosynthetic yields and their final fates.

Before we describe the nucleosynthetic signatures of the oldest intermediate-mass stars we should recall that massive stars are still preferred by many authors as the main, and perhaps the only, genuine `first stars', and thus the first and only polluters of the most primitive Universe. 
All primordial massive star models and, especially, hypernovae (\citealt{nakamura2001b}, \citealt{nomoto2001}, \citealt{umeda2005a}) provide the high $\alpha-$enhancements observed in many EMP stars (item i in section \ref{sec:observations}), and yield relative Fe-peak element abundances in good agreement with many observed EMP stars.
Faint SNe experience extensive fallback of the ejecta and re-accretion onto a central black hole. The part of the ejecta that is not re-accreted (the actual nucleosynthetic yields) is characterised by large [C/Fe] and [Al/Fe] compared to the yields from SNe which do not experience significant fallback  (\citealt{bonifacio2003}; \citealt{lim03}; \citealt{umeda2003}; \citealt{umeda2005b}; \citealt{tominaga2014}). These yields are consistent with the abundances of some observed CEMP-no stars (items d, f and g of section \ref{sec:observations}).

Spinstars or fast rotating massive stars were probably frequent among low-$Z$ objects because of their compactness. As a consequence of enhanced mixing due to rotation, they produce large amounts of primary $^{13}$C, $^{14}$N,
and $^{22}$Ne (\citealt{mey05}; \citealt{mey07}; \citealt{hirschi2007}; \citealt{ekstrom2008}; \citealt{cescutti2013}), and have been proposed as promising candidates to explore the trend of increasing N/O at lower metallicities in EMP stars (item h of section \ref{sec:observations}). Rotating massive star models have even been proposed as sites for the formation of $s$-process elements \citep[see e.g.][and references therein]{frischknecht2016}. 
For a detailed review of yields from massive stars the interested reader is referred to \citet{nomoto2013}.

The possible contribution of an early population of intermediate-mass stars to the chemical evolution of the ancient universe was addressed by \citet{vangioni2011}.
Based on comparisons between theoretical yields and observations, these authors concluded that the influence of intermediate-mass metal-poor stars would probably be restricted to a limited fraction of the total baryon content of the universe. 
However their use of yields (from \citealt{vandenhoek1997}) for
relatively high metallicities of $Z\geq 0.001$ neglects the nucleosynthetic peculiarities of the most metal-poor stars $ Z \lesssim 10^{-6}$, as described later in this section. This suggests that an account of more recent low-$Z$ data is required.
Besides considering the contribution to the baryon inventory, it would be interesting to consider timescales for chemical enrichment by intermediate-mass stars provided by galactic chemical evolution models. However, the lack of 
consistent detailed yields for these intermediate-mass models at the lowest metallicity regimes also limits the assessment of their contribution  which we can derive from chemical evolution models.

The scatter in metal abundances at the lowest [Fe/H] stars mentioned in item a of Section \ref{sec:observations} can be interpreted in terms of differences in the environment where the oldest stars formed. These environments were  primitive gas clouds only polluted by one or a few stars, which might have different masses in different clouds and, therefore, experienced different nucleosynthetic processes (see, e.g., \citealt{bonifacio2003}, \citealt{lim03}). Item b is telling us about the complexity of structure formation in the Milky Way. Items c and j are some of the strongest evidences of our incomplete knowledge of the physics of stars (at the lowest $Z$ regime). We now describe relevant nucleosynthetic sites in low $Z$ low- and intermediate mass stars, and try to explain the remaining items of section \ref{sec:observations}.

\subsection{Dual Flash/C-ingestion nucleosynthesis}

The evolution through core and shell flashes and proton-ingestion was briefly summarised in Sections \ref{sec:prevmixing} and \ref{sec:tpagb}. 
These mixing events occur in extremely metal-poor models of initial mass $\rm M_{ZAMS}\lesssim 4$~\msun, at different locations inside the star and at different evolutionary stages, depending on the initial mass and metallicity. They all involve the entrainment of proton-rich matter into a He-burning convective region.
Stellar models (see, e.g. \citealt{fujimoto2000}, \citealt{Schlattl2002}, \citealt{picardi2004}, \citealt{campbell2008}, \citealt{suda2010}) indicate that dual flashes lead to a significant enrichment of the envelope in carbon and nitrogen. The detailed nucleosynthesis associated with this process was studied by \cite{campbell2010} and \cite{cruz13}. \cite{cristallo2009,cristallo2016} also analysed PIEs at [Fe/H] = -2.85.

As a consequence of a PIE, relatively high amounts of $^{13}$C form and lead to a large release of neutrons via the $^{13}$C($\alpha$,n)$^{16}$O reaction and to the production of heavy $s$-elements like Sr, Ba and Pb. 
Simultaneously, high amounts of ${\rm ^{14}N}$ are produced during these PIEs. This isotope acts as a neutron poison via ${\rm ^{14}N(n,p)^{14}C}$, and may effectively halt $s$-process nucleosynthesis \citep{cruz13}.

Neutron capture nucleosynthesis at the lowest metallicities, although critical,  is still incomplete and part of the reason is due to our limited understanding of the physics of these PIEs. Further investigations using multidimensional hydrodynamical models (for instance, as in \cite{stancliffe2011}, \cite{herwig2011}, and \cite{woodward2015} and references therein) and considering the effects of convective overshooting, extra-mixing and rotationally-induced mixing should be carried out. Observationally, many CEMP stars show  $s$-process enrichment (i.e. they are class CEMP-$s$, see items d and e in section \ref{sec:observations}).
We have seen that a significant number of objects show both $r$- and $s$-enrichment (CEMP-$r/s$) stars (see Section \ref{sec:nucleosynthesis}). This is puzzling because $r$- and $s$- processes are supposed to occur in very different nucleosynthetic sites. The intermediate $i$-process \citep{cowan1977}, occurring at neutron density regimes between the $s$- and the $r$-process might be a key to interpreting CEMP-$r/s$ (see \citealt{abate2016}, and references therein, for different scenarios for the formation of CEMP-$r/s$ stars). A good understanding of the $i$-process, and the interpretation of surface abundances of CEMP-$r/s$ stars probably involves the necessity of 3D-hydrodynamical codes to properly account for the transport of processed matter \citep{dardelet2014}. 
Nevertheless some interesting results concerning $i$-process nucleosynthesis were presented by \cite{Hampel2016}. They performed detailed nucleosynthesis
for high neutron densities characteristic of PIEs in CEMP stars. Although their analysis was not self-consistent, in the sense that it did not involve evolutionary model calculations, these authors found a remarkable agreement between their parametric $i$-process calculations and the abundances of CEMP-$r/s$ stars, even suggesting that they be called CEMP-$i$ stars in future.

\subsection{Nucleosynthesis in models leading to SN~I1/2} \label{sec:ns-sni12}

We have seen in sections \ref{sec:evolution} and \ref{sec:fates} that some intermediate-mass stars ($\rm 4 ~M_\odot \lesssim M_{ZAMS} \lesssim 7~M_\odot$) of initial metallicity $Z_\mathrm{ZAMS}\lesssim 10^{-8}$  experience weak envelope pollution and might end their lives as SNe~I1/2.

In the absence of significant mass ejection prior to the SN explosion, 
and if thermal pulses do not re-ignite  \citep{lau2008},
one expects the yields of these stars to be very similar to those of thermonuclear SNe~Ia \citep{tsujimoto2006} with a  contribution from hot bottom burning nucleosynthesis. 
Explosive nucleosynthesis would lead to large amounts of $^{56}$Ni and other Fe-peak elements, with ratios similar to those of a standard SN~Ia (\citealt{nomoto1984a}, \citealt{nomoto2013}). Nucleosynthesis above the CO core after the SN explosion does not seem likely, because, by analogy with SNe~Ia, the combustion flame is expected to be extinguished before it reaches the H-rich envelope, and thus explosive nucleosynthesis would remain confined to the core. As in SNe~Ia, explosive nucleosynthetic yields of SNe~I1/2 will be significantly affected by the details of the explosion mechanism \citep[see e.g.][and references therein]{mazzali2007}. It is also important to note the presence of high amounts of H from the relatively massive envelope existing at the moment of the explosion would also be present in the SN~I1/2 spectrum, and thus make it more similar to that of type-II SN in this respect.

The relevance of hot bottom burning nucleosynthesis is model dependent. The primordial 5 and 7~\msun{} stars from 
\cite{lau2008} showed a relatively mild hot bottom burning, leading to ${\rm X_{surf}(^{14}N)/X_{surf}(^{12}C)}\sim$ 5 at the end of thermal pulses, whereas the same models computed with overshooting led to 
${\rm X_{surf}(^{14}N)/X_{surf}(^{12}C)} \sim 100$ at the end of calculations \citep{lau2009}.
The surface abundances of the primordial 4~\msun{} model in Figure \ref{fig:tpagb4ms} do not show any effect of hot bottom burning until after the cessation of thermal pulses. However, when this process occurs, it develops as a very hot hot bottom burning. The nucleosynthetic signatures of such extreme hot bottom burning are primarily a large production of He but also \chem{12,13}C, \chem{14}N, and even of some O isotopes.  
Additionally, although no $s$-process elements are dredged up during the AGB phase of these stars, they are produced in the intershell (via $^{22}$Ne neutron source).  
The products processed during pre-supernova evolution could either be expelled in the SN~I1/2 explosion, adding to the ISM inventory of s-process elements, or destroyed during the explosion itself. Detailed calculations should be performed in order to obtain the detailed nucleosynthetic yields.

SN~I1/2 in binary systems have been suggested as possible candidates to explain CEMP-$r/s$ stars (item d of section \ref{sec:observations}) by several authors \citep{zij04,wanajo2006,abate2016} but these progenitors present a number of problems, e.g. population synthesis studies do not reproduce the observed proportion of CEMP-$s$ to CEMP-$r/s$ stars \citep{abate2016}. 
It should also be noted that many authors consider that the SN~I1/2 explosion would destroy the progenitor \citep{nomoto1987}, so the resulting CEMP stars would not be detected as binaries. However, \cite{hansen2016b} showed the existence of single CEMP-s stars and the occurrence of single CEMP-r/s cannot be discarded.

\subsection{Nucleosynthesis in EMP stars undergoing `normal' thermally-pulsing AGB and Super-AGB evolution}

\begin{figure}[t]
\includegraphics[width=1.0\linewidth]{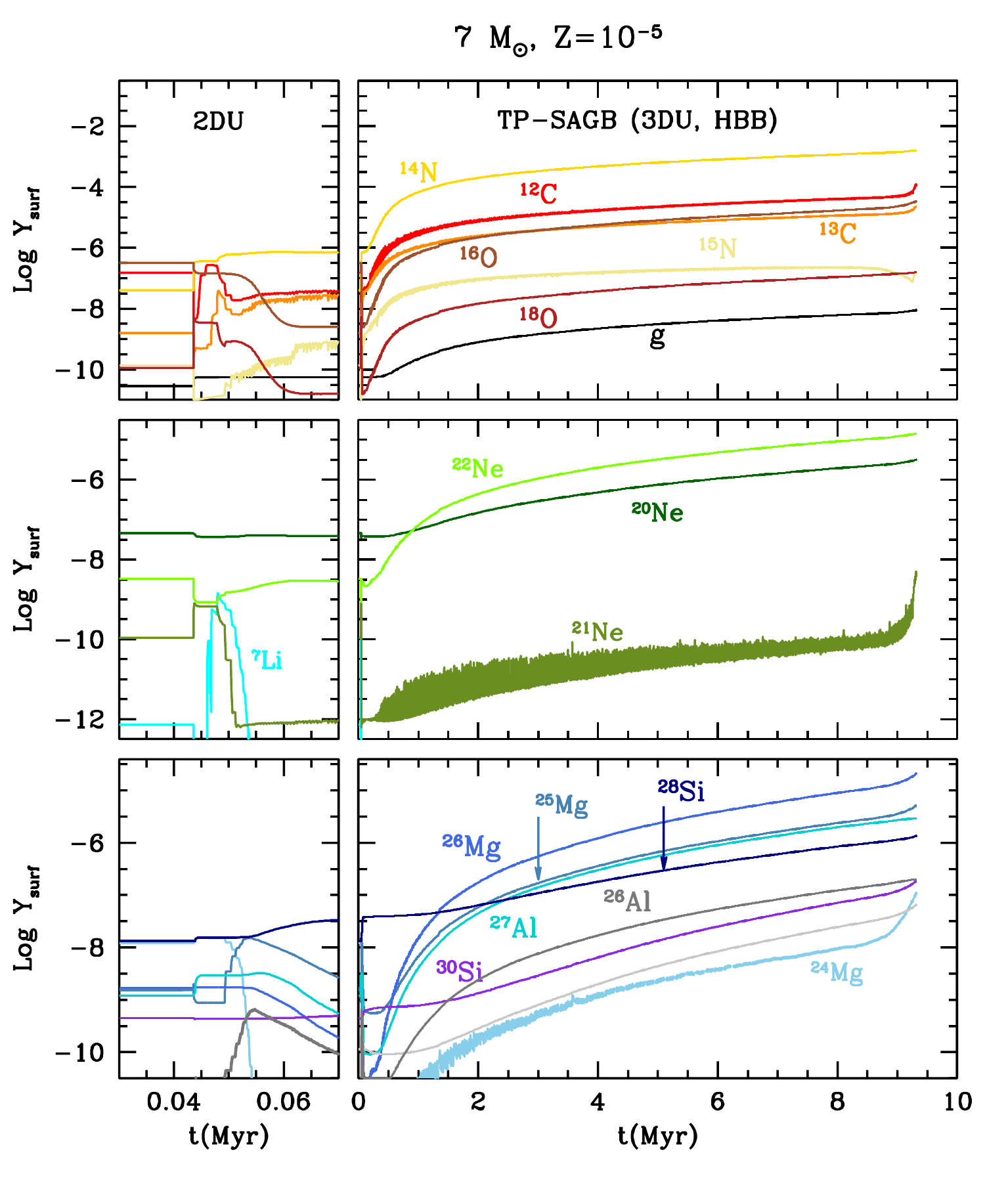}
\caption{Evolution of the surface abundances of some selected isotopes for a 7~\msun{} model with $Z=10^{-5}$ computed with \textsc{monstar} and \textsc{monsoon} (see text for details).}
\label{evoisot}
\end{figure}

Intermediate-mass primordial models which implement some overshooting below the envelope allow for more or less efficient third dredge-up, envelope pollution and stellar winds \citep{chieffi2001,siess2002}. 
The efficiency of the third dredge-up process thus has a strong impact on the yields and the dependence on the stellar mass was studied by \cite{gilpons2013} in their $Z=10^{-5}$ models. These authors, who use the  search for neutrality approach to determine the convective boundaries \citep{frost1996},  obtain high values of the dredge-up parameter $\lambda$\footnote{The $\lambda$ parameter is defined as $\lambda=\frac{\Delta M_\mathrm{dredge}}{\Delta M_\mathrm{core}}$, where $\Delta M_\mathrm{dredge}$ is the H-exhausted core mass dredged-up by the convective envelope after a thermal pulse, and $\Delta M_\mathrm{core}$ is the amount by which the core has grown during the previous interpulse period.} for model stars up to 7~\msun, for which $\lambda=0.78$. This value decreases with the stellar mass ($\lambda=0.48$ for the 8~\msun{} model), and becomes very small ($\lambda=0.05$) for the 9~\msun{} model. Together with a thorough analysis of the evolution, \cite{gilpons2013} presented a limited set of  nucleosynthetic yields for stars between 4 and 9 \msun, including $^1$H, $^4$He, $^{12}$C, $^{14}$N, $^{16}$O, and Z$_{other}$, representing all the isotopes beyond $^{16}$O.

The nucleosynthetic yields of intermediate-mass and massive-stars were computed by \cite{chieffi2001} and  \cite{limongi2000}, respectively. \cite{abia2001} used these existing yields to assess the contribution of intermediate-mass and massive stars to the the pollution of the early intergalactic medium. \cite{campbell2008} also presented yields of primordial and very low metallicity stars in the low- and intermediate-mass range, although only up to 3~\msun.
Primordial star yields in the intermediate-mass range are strongly affected by the unknowns in mass-loss rates and dredge-up efficiency during the thermally-pulsing AGB phase. Therefore, a detailed study of the effects of different input physics, not only for primordial compositions but also up to initial metallicity  $Z = 10^{-5}$, is badly needed. 

Nevertheless, we can attempt to draw some conclusions from the existing literature.
As we may expect from the results for primordial stars of $\rm M_{ZAMS} \gtrsim$ 3~\msun{} which have experienced efficient envelope pollution, the models of 3~\msun{}$\rm \lesssim M_{ZAMS} \lesssim$ 7~\msun{} by \cite{chieffi2001} and \cite{siess2002} show efficient hot bottom burning.  
In general models that experience hot bottom burning
display an increase in their surface abundances of $^4$He and $^{14}$N at the expense of $^{12}$C. 
However, the very high temperature at which hot bottom burning is operating in massive AGB and Super-AGB stars leads to a slight production of $^{12}$C. This is also seen in the more metal-rich Super-AGB stars of \cite{siess2010} that do not experience third dredge-up.
Besides, $^{23}$Na is processed at the expense of $^{22}$Ne, and $^{26}$Al from $^{25}$Mg. The $^7$Li produced during hot bottom burning would be quickly destroyed and thus its contribution to yields would be negligible (\citealt{abia2001} and \citealt{siess2002}). Depending on the efficiency of the third dredge-up, the surface $^{12}$C can be strongly affected \citep[see e.g.][]{doherty2014b}.

\cite{siess2003} analysed $s$-process nucleosynthesis in a primordial 3~\msun{} star. They found that the neutrons released from the $\rm ^{13}C(\alpha,n)^{16}O$ reaction would be captured by isotopes between C and Ne. The heavier species synthesized would then act as seeds to form $s$-process elements. Once transported to the surface by third dredge-up, these stars are expected to display Pb and Bi enhancements  \citep[see also][]{suda2017b}. \cite{cruz13} also computed and analysed s-process nucleosynthesis in 1~\msun{} stars between primordial and $Z=10^{-7}$. They emphasised the effects of input physics uncertainties on their yields.

We now illustrate the detailed nucleosynthesis of $Z=10^{-5}$ models by showing results computed with \textsc{monstar} and  the post-processing nucleosynthesis program \textsc{monsoon}, e.g. \cite{doherty2014a}.  Figure \ref{evoisot} shows a 7~\msun{} model (Gil-Pons et al. 2018, in preparation). The effects of hot bottom burning (the average temperature of the base of the convective envelope during the thermally-pulsing Super-AGB phase is 114$\times 10^6\:K$) can be seen in the increase in $^{14}$N, $^{13}$C and $^{17}$O and, to a lesser extent, of $^{21}$Ne and $^{26}$Mg, together with a decrease of $^{15}$N. The onset of the Mg-Al chains results in the depletion of most $^{24}$Mg and an increase in $^{26}$Al, which at high temperatures captures a proton to give $^{27}$Al \citep{siess2008}, and subsequently $^{28}$Si \citep{ventura2011b}. Some of the effects of hot bottom burning are suppressed by efficient third dredge-up, which replenishes $^{12}$C after each pulse. 
$\alpha-$captures on $^{12}$C in the intershell convective region and subsequent third dredge-up produce surface enhancements in $^{16}$O, $^{20}$Ne, and $^{24}$Mg while $^{28}$Si production is mainly due to proton capture reactions and a leakage from the Mg-Al chain.
It is also important to note the $^{22}$Ne enhancement, because the occurrence of the $\rm ^{22}Ne(\alpha,n)^{25}Mg$ reaction may be an important source of neutrons and, consequently, relevant for $s$-process nucleosynthesis in massive AGB and Super-AGB stars.

It has been reported that stars with $Z_\mathrm{ZAMS}\leq 10^{-4}$ and masses above 8~\msun{} experience high envelope pollution caused by corrosive second dredge-up \citep{gilpons2013,doherty2014b}. The large amount of $^{12}$C dredged-up during this event increases the molecular opacities in the envelope and then drives stellar winds similar to those of a higher $Z$ object. 
These low-$Z$ Super-AGB stars also present very efficient hot bottom burning, but their low third dredge-up efficiency together with the thinness of the intershell regions hampers the possibility of a strong $s$-enhancement in models with $\rm M_{ZAMS} \gtrsim$ 8~\msun. Their nucleosynthesis is similar to that of their slightly lower mass hot bottom burning counterparts. The yields of all the models computed by \cite{gilpons2013} and, in particular, for their 8 and 9~\msun{} models, have [C/Fe] $\geq$ 2. 
If this feature is maintained at the lowest metallicities ($Z < 10^{-5}$), 8 to 9~\msun{} stars of the first (few?) generation(s) would then have the same properties as some CEMP-no stars, making them potential progenitor candidates. 
According to the present IMF this mass range does not account for a significant number of stars, but given that the primitive IMF might be biased to higher masses, their contribution might be relevant.
These models might also help to explain some NEMP stars, described in item h of section \ref{sec:observations}, as polluters of the gas clouds in which they formed.

\cite{meynet2002} investigated the evolution of rotating $Z=10^{-5}$ models and obtained high  $^{12}$C and $^{14}$N surface enrichments in  their intermediate-mass stars. However these authors only computed a few thermal pulses and therefore no complete nucleosynthetic yields were provided.

The most massive Super-AGB stars, which experience a dredge-out process, have been suggested as a site for the formation of neutron-capture elements and, in particular, for the occurrence of the $i$-process (\citealt{petermann2014}, \citealt{doherty2015}, \citealt{jones2016a}). This intriguing hypothesis is still to be demonstrated and carefully analysed, probably requiring 3D hydrodynamical techniques.

The low- and intermediate-mass EMP stars considered in this section are  also likely to have a  binary companion.
Actually binarity has been a key to some of the most successful scenarios to interpret EMP stars (see, e.g. \cite{starkenburg14} and references therein).
If a star undergoing a dual flash, or simply third dredge-up of $s$-process elements, is the primary component (initially the more massive star) of an interacting binary system, then the $s$-process elements synthetized by the primary can be transferred to its companion. If such a companion has a mass $\rm M_{ZAMS}$ about 0.8~\msun{} it can survive to the present day and be detected as a CEMP-$s$ star, as referred to in items d and e. Note that high amounts of C are expected to be dredged-up, together with the $s$-process elements. This binary scenario for the formation of CEMP stars (e.g. \citealt{suda2004}) was in agreement with the radial velocity data of CEMP-$s$ stars, which was consistent with all of them being members of binary systems (\citealt{lucatello2005}, \citealt{starkenburg14}). 
However, updated results of radial-velocity monitoring of CEMP stars show that not all CEMP-s stars are in
binary systems \citep{hansen2016b}, although the percentage of CEMP-s in binaries is still considerably higher than in normal metal-poor stars.

\subsection{Cautionary remarks}\label{sec:caution}

One should be cautious when interpreting EMP abundances using nucleosynthetic yields of model stars.
To begin with, if the observed object is a giant, it may have undergone internal pollution as a consequence of evolutionary processes. Additionally, even dwarf stars may experience mixing processes such as thermohaline mixing \citep{stancliffe2011}, gravitational settling (\citealt{richard2002}, \citealt{macdonald2013}), radiative levitation \citep{matrozis2016}, 
mixing induced by rotation or gravity waves \citep[e.g.][]{talon2008}, or accretion from the ISM (\citealt{yoshii1981}, \citealt{iben1983}, \citealt{komiya2015}).
All these processes may alter surface abundances after accretion from a more evolved companion star, and must be disentangled if we are to understand the stellar nucleosynthesis.

The problem of interpreting the abundances of individual EMP stars is complicated because some of these stars may originate from a second stellar generation.
This second generation probably formed in mini-halos \citep[see e.g.][]{schneider12,chiaki2013,ji2015}), as we think Pop III stars did, in a cloud polluted by gas from a few SN explosions, which was partially retained and partially ejected from the minihalo. Some of the ejected gas could have been re-accreted and then mixed with originally pristine gas and matter from nearby SNe. Therefore nucleosynthetic yield information should be complemented with chemical evolution models which take into account mixing and turbulence \citep{ritter2015}.

\section{SUMMARY AND DISCUSSION}

\subsection{Summary}
The birth, evolution, fate and nucleosynthetic yields of the first generations of stars remain, in many senses, enigmatic. We have seen that the solution to this puzzle is hampered by the specific computational problems which plague the evolution of the most metal-poor stars (such as violent thermonuclear runaways,  thousands of thermal pulses, or unexpected instabilities), by the high sensitivity of results to the details of very uncertain input physics (in particular to opacities, mass-loss rates, convection and mixing, as well as some key nuclear reaction rates), and by the difficulties in obtaining constraints from observational data.

The occurrence of primordial low- and intermediate-mass stars, strongly debated during the last few decades, is supported by recent high resolution 3D-hydrodynamical calculations
of primordial star formation.
In terms of the final fates of intermediate-mass stars, different authors agree (except for the precise mass threshold), that primordial to $Z\sim 10^{-7}$ stars of initial mass $\rm M_{ZAMS}\lesssim$ 4~\msun{} experience efficient mixing episodes 
(\citealt{campbell2008}, \citealt{lau2009}, \citealt{suda2010} and references therein), 
either prior to or during the first pulses of their thermally-pulsing AGB phase. These processes enrich the stellar envelopes in metals, and permit later evolution to take place in a way that is very similar to that of higher $Z$ stars. Thus we expect these stars to form white dwarfs. In the low metallicity range considered in this review, the same fate is expected for stars in the mass range 8~\msun{} $\rm \lesssim M_{ZAMS} \lesssim$ 9.5~\msun. On the other hand, the fate of $Z\lesssim 10^{-7}$ stars between $\sim$ 4 and $\sim$ 7~\msun{} is more intriguing, and whether they end as white dwarfs or SNe strongly depends on the choice of input physics. The use of different algorithms to determine convective boundaries may lead to the occurrence of SNe~I1/2 \citep{gil07a,lau2008}, whereas the inclusion of overshooting would probably lead to the formation of white dwarfs \citep{chieffi2001,siess2002}.
We find that the mass range for EC-SNe is relatively narrow, of the order of $\sim$0.2~\msun{} between $\sim 9.2-9.5$~\msun{} and  $\sim 9.7-9.9$~\msun{} for the $Z=10^{-5}$ and primordial cases, respectively.

The nature, evolution, and fate of models of ancient stars must be tested by comparing nucleosynthetic yields with observations of the most metal-poor objects. 
The sample of metal-poor stars has significantly increased during the last decade, but
the interpretation of the surface abundances remains difficult because of internal mixing processes, potential pollution by the ISM, and because the chemodynamical evolution of their parental clouds is not well understood. 
Many observational features may be reproduced by rotating massive stars \citep{maeder2015} and supernova models (\citealt{umeda2003}, \citealt{tominaga2014}) or by low- and intermediate-mass models in binary stars \citep{suda2004}. Traditionally, CEMP-no stars were interpreted as 
second generation stars formed from a mixture of pristine material and ejecta from massive Pop III stars, while the CEMP-$s$ stars were thought of as the low-mass primordial  (or second generation) companion of an intermediate-mass star that went through its thermally-pulsing AGB phase and then polluted its low-mass partner with $s$-elements. 
We show in this work that primordial intermediate-mass model stars might also help to explain some cases of the heterogeneous CEMP-no group, 
and that massive star models including rotation may account for 
some $s$-process enhancement (\citealt{cescutti2013}, \citealt{frischknecht2016}, and 
\citealt{choplin2017}), and thus for the formation of some CEMP-s stars.
The present classification of observations, albeit useful, might mask the nucleosynthetic contributions of stars over a continuous mass and metallicity range. 

Finally, it is important to note that the relatively restricted sample of observed EMP stars is not the only limitation. An understanding of the existing observational results will probably remain incomplete until modelling the entire evolution of intermediate-mass EMP stars with reasonably precise input physics is possible.

\subsection{Present open questions}
In spite of the wealth of interesting results obtained during the last decades, both from the theoretical and the observational point of view, 
many questions related to EMP stars remain unanswered.
\begin{itemize}
\item[i)] Do low- and intermediate mass stars exist at all $Z$, or is there a critical metallicity below which they cannot form? If such a limit exists, it is important to know if its value is closer to 10$^{-8}$ or to 10$^{-6}$. Stars born with the former metallicity behave similarly to primordial objects and, for instance, might allow the formation of SNe~I1/2, whereas the general behaviour of $Z=10^{-6}$ objects more resembles that of `normal' metallicity stars, at least in terms of their final fates. 

\item[ii)] Did SNe~I1/2 ever explode? If they have existed there might be interesting observational consequences. They would synthesise large amounts of Fe-peak elements, and thus might provide a substantial increase in the injection of Fe-group elements much earlier than that provided by SN~Ia explosions. 
The problem is that early Fe should also be significantly produced in primordial hypernovae, and thus the actual origin of this element in the primitive Universe will not be easy to disentangle, unless additional isotopes of intermediate-mass and heavy metals are considered.  
Stars which are simultaneously very old and relatively metal-rich might be detected by using asteroseismology techniques applied to Galactic archeology, as proposed by \cite{miglio2013}. 
Additionally, \citet{bergemann2016} presented a new method to determine ages of red-giant stars, for [Fe/H]\,$\leq -2$. 
However, it is critical to highlight that the huge uncertainties in models of EMP stars may considerably complicate age determinations. A fruitful application of either age-determination method and, eventually, the assessment of the contribution of SNe~I1/2 to the chemical evolution of the Universe should, in any case, use detailed nucleosynthetic yields of models leading to these supernovae.
In relation to possible descendants of SNI1/2 it is interesting to consider stars from the 
Galactic bulge. According to cosmological models (see, for instance, \citealt{white2000}, and \citealt{tumlinson2010}) the Bulge should host the 
oldest stars in the Galaxy. However, observations show that the average metallicity of bulge stars is higher than those from the Halo. Besides, metal-poor stars detected in the Bulge present intriguing peculiarities, such as the absence of C enhancement, and large $\alpha$-element scatter \citep{howes2014,howes2016}. The interpretation of these peculiarities will shed light on our understanding of the oldest stars and, perhaps, on SNI1/2.
The latter explosions might actually appear in the high redshift transient records of new generation telescopes. However, given the relatively low brightness expected for SN~I1/2, a more promising possibility might be to look for them among the supernovae discovered in gravitational lenses (\citealt{quimby2013}, \citealt{kelly2015}; \citealt{goobar2017}), as brightness magnifications of up to $\times$2000 have been observed \citep{kelly2018}. While the supernova brightness could be affected by microlensing due to individual objects in the lensing galaxy \citep{dobler2006}, their spectra would be unaffected, and could become an effective way to classify the observed supernovae.

\item[iii)] What are the roles of overshooting, extra-mixing processes, and rotation, in the evolution of EMP stars?
This question is related to item ii, as we have seen that the inclusion of overshooting may avoid the formation of SNe~I1/2. Additional mixing induced by rotation might lead to effects similar to those of overshooting.

\item[iv)] If low-mass ($\rm M_{ZAMS}\lesssim 0.8$~\msun{}) primordial stars ever formed, could they be unambiguously detected? The possibility that Fe-deprived objects might remain as such is another matter of debate.
\cite{frebel2009} performed kinematical analysis on extensive samples of metal-poor stars, and concluded that ISM pollution was practically negligible. If this is the case, the absence of detection of Fe-deprived objects would be a direct consequence of the fact that they do not exist, at least for initial masses $\rm M_{ZAMS}\lesssim 0.8$~\msun. 
\cite{tanaka2017} and \cite{suzuki2018} performed magnetohydrodynamical simulations for stellar winds driven by Alfv\'en waves, and also determined 
that ISM accretion on primordial low-mass stars should be negligible.
On the other hand,  
\cite{komiya2015} concluded, on the basis of chemical evolution studies
that accretion from the ISM might lead to primordial envelope pollution values as high as [Fe/H] $\sim -5$. 
\item[v)] Could CEMP-no stars form from low- and intermediate mass objects? CEMP-no stars are traditionally assumed to 
have formed from a previous generation of massive stars from which they inherited their chemical peculiarities, but doubts have been cast on this hypothesis. Considering the continuity of the [Ba/C] distribution as a function of [Fe/H] in CEMP-$s$ and CEMP-no stars, \cite{abate2015a} and \cite{suda2017b} suggested that CEMP-$s$ and (some) CEMP-no objects might have a common origin involving binarity. 
Observational studies that analysed the binary fraction of different subclasses of CEMP stars support this hypothesis \citep{starkenburg14,hansen2016a}. 
Similar ideas are discussed in terms of carbon abundances in CEMP-no stars. \cite{bonifacio2015} define two groups of CEMP stars, namely high- and low-carbon band stars. They
insist that high-carbon band stars, consisting of almost all the CEMP-s stars and some CEMP-no
stars, are in binaries. On the other hand, the classification of CEMP stars by \cite{yoon2016} leads to a different conclusion. They consider that the carbon enhancement of CEMP-no stars
is intrinsic, and due to the enrichment of their natal clouds by high-mass progenitor stars.
\item[vi)] Could CEMP-$s$ stars be the offspring of massive stars?
The standard scenario for the formation of CEMP-$s$ stars involves a binary. 
However, recent studies \citep{hansen2016b} revealed the 
existence of isolated CEMP-s stars. The fact that massive star models with different rotation rates can reproduce the observed [Sr/Ba] spread in CEMP stars \citep{cescutti2013,frischknecht2016} provides additional support to this scenario which was recently re-investigated by \cite{choplin2017}.

\end{itemize}

\subsection{Future topics of research}
 
Below we discuss some bottlenecks in our understanding of EMP stars, and also areas that may provide promising avenues for further research.

\begin{itemize}
\item[i)] As is always the case, a better understanding of convection and, in particular, of convective boundaries, is a significant barrier to more reliable models. 
When dealing with stars at the most metal-poor regimes we have little insights into the how to model convection and its borders. We are forced to extrapolate or adapt the existing observational and theoretical information from higher $Z$- objects, and we must be aware of the possibility (and high probability) of introducing substantial errors. 
In spite of these uncertainties there is a reasonable consensus on the evolution and fates of the less massive intermediate-mass objects at the lowest $Z$.
On the other hand our knowledge of the final fates of the most metal poor stars ($Z \lesssim 10^{-7}$) of masses between $\sim$ 4~\msun{} and 8~\msun{} is very poorly constrained. 
Work is proceeding to improve the physics on the treatment of convection and convective boundaries beyond the MLT  \citep[see e.g.][]{arnett2015,campbell2016,arnett2017}.
\item[ii)] A better understanding of low-temperature opacities and mass-loss rates is crucial. Recent improvements in opacity tables by \cite{lederer2009} and \cite{marigo2009} have been implemented in models and their important consequences in terms of stellar wind enhancements have been reported, for instance, in \cite{constantino2014}. The effects of dust in low-temperature opacities might be even more significant \citep{tashibu2017}. Intermediate-mass models with compositions from primordial to $Z=10^{-7}$ should be constructed considering these effects, although the high effective temperature and almost pristine composition of these stars suggest that their evolution would be less sensitive to these changes.

\item[iii)] 
The phenomenon of thermal pulses ceasing and then re-starting is not understood, and is ripe for investigation. We need a consistent set of calculations with different `reasonable' input physics for these models.
The envelope pollution and increase of mass-loss rates associated with the re-onset of thermal pulses might eventually hamper the formation of SNe~I1/2. 

\item[iv)] Many CEMP-s, and some CEMP-no stars have a binary companion. Addressing the problem of their evolution, including mass transfer via wind accretion, should also be a priority (\citealt{bisterzo2011} and \citealt{abate2015b} ).

\item[v)] Improvement in our knowledge of the former issues will help us to obtain
better evolutionary models and nucleosynthetic yields, including full n-capture nucleosynthesis.
Ultimately we want to combine these yields with sophisticated chemical evolution models, in order to get a more realistic approach to the interpretation of EMP abundances (e.g. \citealt{ritter2015}, \citealt{hirai2018}). Dwarf galaxies seem to be promising tools because their formation history is not as complicated as that of the Milky Way. 

\end{itemize}

The current revolution in stellar spectroscopy is changing the landscape. The development of very large telescopes,  enormous surveys and machine learning is driving this revolution. These will allow us to get further information from medium resolution data, so that dwarf galaxies can be analysed \citep{kirby2015}. 
\cite{komiya2016} proposed that Pop III stars freed from their massive companions and undergoing a SN explosion could be detected by large-scale 
giant surveys in the outskirts of the Milky Way.
\cite{magg2018} also calculated the probability of finding Pop III survivors. Their results were compatible with the absence of detection in the Milky Way, but yielded somewhat more promising results for its dwarf satellites. However only giants are expected to be observed in them, which reduces the detection probability.   

The faintness of ancient stars is indeed a challenge for their detection. 
However, if the end of the lives of some of these stars is marked by SN~I1/2 explosions, their luminosity might allow detection  with new generation telescopes such as the James Webb Space Telescope (see \citealt{desouza2014}, and references therein). Detecting and identifying SN~I1/2 explosions would provide us with key information about the primordial IMF and the evolution of the most ancient stars.

\begin{acknowledgements}
The authors thank George Angelou, Takuma Suda, and the anonymous referees for their useful comments.
This work was supported by Spanish MINECO grant AYA2015-71091-P.
This work was supported in part by the National Science Foundation under Grant No. PHY-1430152 (JINA Center for the Evolution of the Elements). CD acknowledges support from the Lendulet-2014 Programme of the Hungarian Academy of Sciences. SWC acknowledges federal funding from the Australian Research Council
though the Future Fellowship grant entitled “Where are the Convective
Boundaries in Stars?” (FT160100046)
\end{acknowledgements}

\bibliographystyle{pasa-mnras}

\bibliography{pilar}

\end{document}